\documentclass[preprint,12pt]{elsarticle}

\usepackage{amssymb,amsmath,amsthm}
\usepackage{graphicx}
\usepackage{amsmath,amsfonts,amssymb,lscape,setspace}
\usepackage{amssymb}
\usepackage{lscape}
\usepackage{setspace}
\usepackage{color}
\usepackage{lineno,hyperref,bm,subfigure,amsmath,amssymb,multirow}
\usepackage{graphicx}

\usepackage{natbib}
\journal{}

\begin{document}

\begin{frontmatter}



\title{First-principle study on compensated half metallic double perovskite compound $Ba_2PrVO_6$}

\author[mymainaddress]{Qiang Gao}
\author[mymainaddress]{Lei Li}
\author[mymainaddress]{Huan-Huan Xie}
%%\author[mymainaddress]{Gang Lei}
\author[mymainaddress]{Jian-Bo Deng}

\author[mymainaddress]{Xian-Ru Hu\corref{mycorrespondingauthor}}
\cortext[mycorrespondingauthor]{Corresponding author}
\ead{huxianru@lzu.edu.cn}

\address[mymainaddress]{School of Physical Science and Technology, Lanzhou University,
 Lanzhou 730000, People's Republic of China}

\begin{abstract}
The first-principle study on double perovskite compound Ba$_2$PrVO$_6$ has been presented in this paper. By analysing band structures and integrated density of states, it is found that Ba$_2$PrVO$_6$ is ferromagnetic metallic within LSDA and compensated half metallic whithin LSDA+U. According to the total and partial density of states, the anti-ferromagnetism of Ba$_2$PrVO$_6$ is originated from the spin down state Pr$^{4+}$ ($4f_{-2}$) and the spin up hybridized state between the partially filled $t_{2g}$ state of V and the partially filled triple-degeneration state $4f_{-1}$, $4f_{+1}$, $4f_{+3}$  of Pr. We have investigated the magnetic evaluation of Ba$_2$PrVO$_6$ through fixed spin moment calculations, and the results indicate that the compensated half-metallic state is the ground state. The thermodynamic properties of Ba$_2$PrVO$_6$ are also studied by employing the quasi-harmonic Debye model. 

\end{abstract}

\begin{keyword}
Anti-ferromagnetism \sep Electronic structure \sep Thermodynamic  properties \sep Magnetic properties \sep Double perovskite

\end{keyword}

\end{frontmatter}


\section{Introduction}

Spin-polarized electron injection from a ferromagnetic metal to a diffusive semiconductor has become a promising field of spintronics during the past decades \cite{1, 2}. Since the prediction of half-metallic (HM) magnetic Heusler alloy NiMnSb in 1983 \cite{3}, the HM materials have been considered as very significant candidates of spin-injector materials due to the high spin polarization (100\%) around Fermi level. As the potential meaning of HM materials in spintronics, much effort has been taken to investigate the HM properties and materials in both theoretical \cite{4, 5} and experimental researches \cite{6, 7}.  

The HM materials behave metallic character in one spin channel, while insulating or semiconducting character in the opposite spin-channel. Thus they can generate spin-polarized currents with no external operation, and are useful for spintronics application. However, as reported in some Refs. \cite{8, 9}, there is a drawback for many HM materials that the spin-polarized currents may be hampered by stray fields which stabilize magnetic domains. Because of its zero net total magnetic moment per unit cell, HM anti-ferromagnet which is a subclass of HM materials, is seen as one of the promising material to avoid the stray fields. A HM anti-ferromagnet has not only one conductive channel and one insulative or semiconductive channel, but also zero net moment in a unit cell \cite{10}. Different from common anti-ferromagnet, the magnetic moments of two different magnetic ions in HM anti-ferromagnet do compensate each other. Thus, it is more properly called a compensated half metal (CHM) \cite{11}. CHM has attracted extreme interest to spintronics and is anticipated to generate a single-spin superconductor due to the zero net magnetic moment \cite{12}. For a long time, no accurate CHM has yet been observed by any experiment. However, the exploration of CHM has never stopped \cite{13, 14} due to the excellent properties. Very recently, Claudia Felser and his collaborators  report the realisation of CHM in experiment \cite{15-a}. 

Among half metals, the complex oxides with double perovskite (DP) structure have attracted much attention to researchers on account of their interesting properties. The DP structure oxides have a unit cell twice the size of the regular perovskite structure with a formula of $A_2BB'O_6$. $Sr_2FeMoO_6$ \cite{15} is the first example of DP structure complex oxide reported to be a HM ferromagnet by means of band-structure calculations. More interesting, in 1998 Pickett \cite{16} predicted DP compound La$_2$VMn$O_6$ to be a CHM. Since then, DP oxides have been regarded as the most promising candidates for CHM. So in the past years DP compounds have been investigated widely and sufficiently. For example, Uehara in Ref. \cite{17} reported that DP compounds $CaLaVMoO_6$ and $SrLaVMoO_6$ are nearly to be CHMs by experiment. DP $Ba_2VTO_6$ (T=Nb, Mo) \cite{18} are not HM ferromagnets within local spin density approximation (LSDA), but HM ferromagnets under LSDA+U (considering the Coulomb correlations). DP $La_2CrFeO_6$ is predicted to be a HM ferrimagnet by both experiment \cite{19} and density function theory (DFT) calculations \cite{13}. With LSDA calculations, Lee reported DP $Sr_2CrFeO_6$ to be a simple metallic ferrimagnet in Ref. \cite{20}. However, including Coulomb correlations, DP $Sr_2CrFeO_6$ is predicted to be a CHM \cite{13}, which is also confirmed by fixed spin moment calculations. Hence, the Coulomb correlations may probably influence the DFT calculations of DP oxide when calculated compounds consist of transition metals.

In 2008 \cite{21}, the DP compounds $Ba_2A^{II}UO_6$ ($A^{II}$= Mn, Fe, Co, Ni, Cu, Zn, Cd, Pb) have been synthesized and investigated. These compounds are consisted of rare earth and transition metal elements. Motivated by the above, we have investigated a series of DP compounds $Ba_2ReXO_6$ (Re=Ce,Pr; X=V,Cr,Mn,Fe,Co), which are also consisted of rare earth and transition metal elements. The previous investigations \cite{13, 16} indicate that the DP compounds consisted of two transition metal (or rare earth) elements may probably be a CHM. And indeed, among $Ba_2ReXO_6$ (Re=Ce,Pr; X=V, Cr, Mn, Fe, Co), we find that Ba$_2$PrVO$_6$ is a CHM wihin LSDA+U. In this paper, we will study the band structures, origin of anti-ferromagnetism and magnetic evaluation of Ba$_2$PrVO$_6$. The results indicate that the anti-ferromagnetism of Ba$_2$PrVO$_6$ is originated from the spin down state Pr$^{4+}$ ($4f_{-2}$) and the spin up hybridized state between the partially filled $t_{2g}$ state of V and the partially filled triple-degeneration state $4f_{-1}$, $4f_{+1}$, $4f_{+3}$  of Pr, and the compensated half-metallic state is the ground state. We also study the thermodynamic properties of the compound by employing the quasi-harmonic Debye model.

\section{Structure and calculation method}
The crystal structures are shown in Figure 1, which are synthesized by fixed Pr (4f) and substituted V (3d) site. The cations are treated as $Pr^{4+}$ and $V^{4+}$. In theory, the crystal structure of double perovskite (DP) compound Ba$_2$PrVO$_6$ can be determined by the value of tolerance factor ($t_f$). $t_f$ is defined by
\begin{equation}
\label{}
t_{f}=\frac{r_{Ba}+r_O}{\sqrt{2}(\frac{r_{Pr}+r_V}{2}+r_O)},
\end{equation}
where $r_{Ba}$, $r_{Pr}$, $r_{V}$ and $r_{O}$ are the ionic radius of Ba, Pr, V and O atoms, respectively. $t_f$ is obtained by using SPuDS (Structure Prediction Diagnostic Software) code \cite{22, 23}. We calculate the DP compound by employing the formula $A_2MM'O_6$ of the NaCl M-cation ordered perovskite with space group of Fm3m (NO. 225). In the cubic DP structure, $t_f\approx 1$. When $t_f< 1$, the bond angle of superexchange $M-O-M'$ is not the ideal, $180^\circ$. In consequence, the ideal cubic becomes rhombohedral or orthorhombic. And if $t_f>1$, a hexagonal structure is formed. In our calculation, $t_f=1.0019$,  which is around $t_f\approx1$. Hence, the compound has a cubic crystal structure (shown in Figure 1) with space group Fm3m  indeed. In addition, we can also get the calculated lattice constant is 8.322 \AA. Ba, Pr, V and O atoms occupy 8c (0.25, 0.25, 0.25), 4a (0, 0, 0), 4b (0.5, 0.5, 0.5) and 24e (0.2324, 0, 0), respectively. So the following calculations are based on the obtained crystal structure information.

The present calculations are carried out with LSDA and LSDA+U approaches implemented in the first-principle full-potential local-orbital (FPLO) minimum-basis method code \cite{24,25}. The Perdew Wang 92 potential \cite{26} is used for the LSDA and LSDA+U calculations. In the LSDA+U calculations, there are two popular double-counting schemes, the so called  around mean field \cite{27} and atomic limit, which showing similar results for the DP compounds. We only show the around mean field calculation results. According to the various investigations \cite{29, 30}, we have chosen the on-site Coulomb repulsion U of 7eV and 4eV for Pr and V, respectively. The Hund's integral is set as J=0.95 eV for $4f$ states of Pr, and J=0.87 eV for $3d$ states of V, in which an effective $U_{eff}=U-J$ incorporates the on-site Coulomb (U) and the exchange interaction (J). For the Brillouin zone integration, we use the $k$ meshes of $20\times20\times20$ for all calculations. For a self-consistent field iteration, the convergence criterion is set to both the density ($10^{-6}$ in code specific units) and the total energy ($10^{-8}$ hartree).

\section{Results and discussions}
\subsection{Band structures and number of states for  Ba$_2$PrVO$_6$ within LSDA and LSDA+U}
We have chosen U from 5 eV to 10 eV for Pr and from 1 eV to 7 eV for V. The calculations give consistent results. So in the following, we just choose U=7 eV and U=4 eV for Pr and V, respectively. Within LSDA calculation, Ba$_2$PrVO$_6$ is ferromagnetic metallic. However within LSDA+U calculation, Ba$_2$PrVO$_6$ is found to be a CHM. The band structures within LSDA and LSDA+U calculations are shown in Figure 2. In order to investigate the magnetic properties, we show the integrated density of states (number of states, NOS) in Figure 3.

Firstly, we will discuss the results of LSDA calculations. As can be seen in Figure 2 (a), within LSDA calculations, it is obvious that the spin up bands get through Fermi level, which is a typical character of metallic property. The spin down bands get Fermi level slightly when amplifying the band structure. In order to investigate the property of Ba$_2$PrVO$_6$ within LSDA clearly and deeply, we demonstrate Figure 3 (a). There is no platform in either spin up or spin down direction around Fermi level. We can see that the NOSs show a variation tendency to the Energy around Fermi level in both spin-up and spin-down channels. So, both spin-up and spin-down channels show metallic character. In addition, the NOSs of spin up and spin down are 47.24 and 46.50 at Fermi level, respectively. So the total magnetic moment of Ba$_2$PrVO$_6$ is 0.74 $\mu_B$. Hence, Ba$_2$PrVO$_6$ is not half-metallic ferromagnetic, but ferromagnetic metallic within LSDA.

Secondly, we discuss the results of LSDA+U calculations. As can be seen in Figure 2 (b), within LSDA+U calculations, for Ba$_2$PrVO$_6$, the spin up bands get through Fermi level, showing a metal character. However, there is a band gap in the spin down bands, behaving a semiconductor character. As a result, the spin  polarization is 100\%. In addition, as shown in Table 1, the total magnetic moment is zero for Ba$_2$PrVO$_6$. Hence Ba$_2$PrVO$_6$ is a CHM within LSDA+U calculations. The same as previous, let us focus on the integrated DOS in order to understand the magnetic properties of the DP compound Ba$_2$PrVO$_6$. As can be seen in Figure 3 (b), there is a platform around Fermi level in the spin down integrated DOS. As for the spin up direction, there is no platform around Fermi level. The integrated DOS in spin-up direction shows a variation tendency to the Energy around Fermi level. Taking account into the former aspects, we can get the conclusion that Ba$_2$PrVO$_6$ is half-metallic. In addition, at Fermi level the NOSs of spin down and spin up are 47 and 47, which are the same values. So the net total magnetic moment is zero. Hence Ba$_2$PrVO$_6$ is a CHM within LSDA+U. 

\subsection{The origin of anti-ferromagnetism for Ba$_2$PrVO$_6$}
In this section, we will study the origin of anti-ferromagnetism for\\ Ba$_2$PrVO$_6$. The total and partial density of states (DOS) within LSDA+U for Ba$_2$PrVO$_6$ are presented in Figure 4. The triple-degeneration state of $4f_{-1}$, $4f_{+1}$ and $4f_{+3}$ for Pr is named as $\alpha$ state in our paper. According to the population analysis of electrons, the valencies of both V and Pr are 4+. And the $V^{4+}$ and $Pr^{4+}$ cations have the configurations of $V^{4+}$ ($3d^14s^0$) and $Pr^{4+}$ ($4f^16s^0$), respectively. 

As can be seen in Figure 4, in the spin-up channel, most of the partially filled 6O$-2p$ states lie beyond Fermi level and only a few of them are below Fermi level. While in the spin-down channel, the corresponding 6O$-2p$ states are above Fermi level. So the magnetic moment originated from 6O$-2p$ states is very small (0.19 $\mu_B$).  In the spin-down channel, the strong local state ($4f_{-2}$) is at about -4.8 eV, while the corresponding strong local state is at about 2.5 eV in the spin-up channel. One electron occupies at the strong local state ($4f_{-2}$) in the spin-down channel, which providing a negative magnetic moment. Around Fermi level, the partially filled $t_{2g}$ state of V and the partially filled $\alpha$ state of Pr hybrid with each other in the spin-up channel. One electron occupies at the partially filled hybridized state in the spin-up channel, which providing most of the positive magnetic moment. It is obvious that the anti-ferromagnetism of Ba$_2$PrVO$_6$ is originated from the anti-aligned spin states between ($4f_{-2}$) state and hybridized $t_{2g}-\alpha$ state.

\subsection{Magnetic evaluation of Ba$_2$PrVO$_6$}
As described in Ref.\cite{13}, the CHM state may not be the ground state. In this section, the magnetic evaluation of Ba$_2$PrVO$_6$ is studied by fixed spin moment (FSM) calculations within LSDA+U, and the results are shown in Figure 5. We can obtain the solutions: ferrimagnetic (FI) state below $M_{tot}\approx1.7 \mu_B$ and ferromagnetic (FM) state above $M_{tot}\approx3.5 \mu_B$, while both states coexist in the region from  $M_{tot}\approx1.7 \mu_B$ to $M_{tot}\approx3.5 \mu_B$. 

Let us focus on the FI state at first. As $M_{tot}$ is increased from the CHM state ($M_{tot}$=0), the energy increases monotonically. Besides, considering the symmetry between positive and negative magnetic moment, we can get the $M_{tot}$=0 is a local minimum point of energy to magnetic moment. In FI state, the energy increases almost linearly to the increasing of magnetic moment. $M_{Pr}$ keeps positive value and decreases gradually, while $M_{V}$ keeps negative value and also decreases gradually. So $\frac{M_{V}}{M_{Pr}}$ decreases gradually and keeps below -1. However, beyond $M_{tot}\approx2.5\mu_B$, $M_{Pr}$ changes violently with the increasing of $M_{tot}$, and quickly reaches to zero. Hence, around $M_{tot}\approx2.5\mu_B$, $\frac{M_{V}}{M_{Pr}}$ changes violently to $M_{tot}$, leading to non-analytic behaviour in the FSM curve in this range. The discontinuous trend means phase transition may occur around $M_{tot}\approx2.5\mu_B$. 

Next, we will focus on FM state. As $M_{tot}$ is increased from $M_{tot}=1.7\mu_B$, the energy decreases monotonically and reaches a local minimum at $M_{tot}=2\mu_B$, meaning another local stable state occurs. However, this FM state has a higher energy by 1.28 eV than the CHM state. Hence, the CHM state is the ground state. Below $M_{tot}\approx2\mu_B$, $\frac{M_{Pr}}{M_{V}}$ changes violently to magnetic moment, leading to non-analytic behaviour in the FSM curve in this range. It means the phase transition occurs below $M_{tot}\approx2\mu_B$. In the FM state, $\frac{M_{Pr}}{M_{V}}$ decreases with the increasing of $M_{tot}$, and finally tends to 2. The stable FM state ($M_{tot}=2\mu_B$) is half-metallic ferromagnetic (HMF). The magnetic moments for Ba, V, Pr and 6O are 0.01 $\mu_B$, 0.65 $\mu_B$, 1.43 $\mu_B$ and -0.09 $\mu_B$, respectively. The magnetic moments of Pr and V are parallel now. 

In a word, we have studied the magnetic evaluation of Ba$_2$PrVO$_6$ within LSDA+U, and found that the CHM state is the ground state.

\subsection{Thermodynamic properties}
\subsubsection{Thermodynamic method}
Study of the thermodynamic of materials is very important to the extension of our knowledge about their special behaviours under some external conditions, which can help us in the application of materials. So in the following, we will investigate the thermodynamic properties of Ba$_2$PrVO$_6$. In order to study the temperature and pressure dependences of thermodynamic properties for the crystal,  we have employed the quasi-harmonic Debye model as a simple way to take into account the vibrational motion of the lattice. Here the quasi-harmonic Debye model is implemented in the pseudo code GIBBS \cite{33}, in which the isotropic approximation is considered. As described in Ref \cite{33}, only with a series of (V, E(V)) points around the equilibrium geometry can the code GIBBS calculate out many thermodynamic properties. The non-equilibrium Gibbs energy $G^*(V;P,T)$ is expressed by
\begin{equation}
\label{}
G^*(V; P, T)=E(V)+PV+A_{vib}[\theta(V); T],
\end{equation}
where E(V), PV, $A_{vib}$ and $\theta(V)$ are the total energy per unit cell, constant hydrostatic pressure condition, vibrational term and Debye temperature, respectively. The vibrational term $A_{vib}$ can be written using Debye model of phonon density of states as:
\begin{equation}
 \label{}
A_{vib}(V; P, T)=Nk_BT[\frac{9\theta}{8T}+3ln(1-e^{-\frac{\theta}{T}})-D(\frac{\theta}{T})],
\end{equation}
where $D(\frac{\theta}{T})$, N and $k_B$ are the Debye integral, number of atoms in a chemical formula and Boltzmann constant, respectively. Using the quasi-harmonic Debye model in Slater formulation, the Debye temperature $\theta_D$ is calculated from the static bulk moduli $B_S(V)$, while Poisson ratio ($\nu$) is assumed to be volume independent and equal to $\frac{1}{4}$, corresponding to a Cauchy solid.  Considering the assumption of isotropic conditions, as a function of $\nu$ and V, $\theta_D$ can be given by \cite{34}
\begin{equation}
 \label{}
\theta_D=\frac{h}{k_B}[6\pi^2V^{\frac{1}{2}}N]^{\frac{1}{3}}f(\nu)\sqrt{\frac{B_S}{M}},
\end{equation}
where M is molecular mass per formula unit. $f(\nu)$ is expressed as
\begin{equation}
 \label{}
f(\nu)=\{3[2(\frac{2}{3}\frac{1+\nu}{1-2\nu})^{\frac{3}{2}}+(\frac{1}{3}\frac{1+\nu}{1-\nu})^{\frac{3}{2}}]^{-1}\}^{\frac{1}{3}}.
\end{equation}

The static bulk modulus can be computed by the curvature of the $E(V)$ function:
\begin{equation}
\label{}
B_S\cong B_{static}=V(\frac{d^2E(V)}{dV^2}).
\end{equation}

The curvature of the $E(V)$ function changes with volume, increasing sharply as the compression of crystal, and decreasing gently as the expansion of crystal. This asymmetry between the curvature at the left and right of the equilibrium volume is originated from the volume dependence of $\theta_D$ and is the main reason behind the capability of the quasi-harmonic Debye model to predict the low temperature behaviour of crystal, including $V(T)$ dependency. Accordingly, the non-equilibrium Gibbs function $G^*(V; P, T)$ as a function of V, P and T can be minimized with respect to V:
\begin{equation}
\label{}
[\frac{\partial{G^*(V; P, T)}}{\partial{V}}]_{P,T}=0.
\end{equation}

By solving the above equation, we can obtain the thermal equation of state (EOS) $V(P,T)$. The heat capacity at constant volume $C_V$, isothermal bulk modulus $B_T$, and  thermal expansion coefficient $\alpha$ can be respectively expressed by:
\begin{equation}
\label{}
C_V=C_V(V(P,T),T)=(\frac{\partial G^*}{\partial T})_V=3Nk_B[4D(\frac{\theta}{T})-\frac{3\frac{\theta}{T}}{e^{\frac{\theta}{T}}-1})],
\end{equation}

\begin{equation}
\label{}
B_T(P,V)=V(\frac{\partial ^2G^*(V;P,T)}{\partial V^2})_{P,T},
\end{equation}

\begin{equation}
\label{}
\alpha=\frac{\gamma C_V}{B_TV},
\end{equation}
where $\gamma$ is the Gr$\ddot{u}$neisen parameter. And $\gamma$ is defined by:
\begin{equation}
\label{}
\gamma=\frac{dln\theta(V)}{dlnV}.
\end{equation}

Through the quasi-harmonic Debye model described above, we have calculated the thermodynamic properties of Ba$_2$PrVO$_6$ at the temperature range from 0 to 500 K and in the pressure range from 0 to 20 GPa, where the quasi-harmonic model keeps valid enough.  

\subsubsection{Thermodynamic results and discussions}
The relationship plots of bulk modulus (B) to pressure (P) at different temperature (T) are shown in Figure 6. As can be seen, B increases linearly with P at a given T. In addition, in our calculated range, the difference of B for various temperature at a given pressure is such small that the plots of 0K, 300K and 500K almost seem to overlapping. Heat capacity is also an important thermodynamic parameter. Through heat capacity $C_V$, we can get the information about lattice vibrations, energy band structures, electron density of states, phase transition of solid and so on. The heat capacity $C_V$ of Ba$_2$PrVO$_6$ as a function of T with P at 0 GPa, 5 GPa, 10 GPa, 15 GPa, 20 GPa and as a function of P with T at 100 K, 200 K, 300 K, 400 K, 500 K are shown in Figure 7 (a) and (b), respectively. As can be seen from Figure 7 (a), under some given pressures the $C_V$ of Ba$_2$PrVO$_6$ is proportional to $T^3$ at low temperature. In addition, $C_V$ increases monotonically with temperature and tends to a constant value at high temperature, which is the Petit and Dulong limit \cite{35} and a common phenomenon of all solids. From Figure 7 (b), we can find that $C_V$ decreases slowly and almost linearly with pressure at given temperature and $C_V$ is almost independent of pressure at high temperature. Hence, we can get the conclusion that the effects of T and P on $C_V$ are opposite, and the influence of the former is more significant than that of the later for Ba$_2$PrVO$_6$. 

The Debye temperature ($\theta_D$) is a critical fundamental parameter closely related to many physical properties, for example specific heat and melting temperature. The Debye temperature $\theta_D$ of Ba$_2$PrVO$_6$ as a function T with P at 0 GPa, 5 GPa, 10 GPa, 15 GPa, 20 GPa and as a function of P with T at 100 K, 200 K, 300 K, 400 K, 500 K are shown in Figure 7 (c) and (d), respectively. As can be seen form Figure 7 (c), $\theta_D$ is slightly linearly decreasing with the increasing of T. The $\theta_D$ at pressure of 0 GPa, 5 GPa, 10 GPa, 15 GPa and 20 GPa is reduced by 2.38\%, 2.15\%, 1.92\%, 1.69\% and 1.52\% when T changes from 0 K to 500 K, respectively. In addition, as shown in Figure 7(d), $\theta_D$ increases quickly but non-linearly with the increasing of P. The $\theta_D$ at 100 K, 200 K, 300 K, 400 K and 500 K are increased by 22.32\%, 22.54\%, 22.82\%, 23.09\% and 23.34\% when P changes from 0 GPa to 20 GPa, respectively. Hence the effects of T and P are also opposite, and the influence of the latter is much more significant than that of the former for Ba$_2$PrVO$_6$. As shown in Figure 7(d), with the increasing of P, $\theta_D$ increases non-linearly and its slope becomes less, revealing the vibration frequency of lattice is changing under pressure. 

The thermal expansion ($\alpha$) of Ba$_2$PrVO$_6$ as a function of T with P at 0 GPa, 5 GPa, 10 GPa, 15 GPa, 20 GPa and as a function of P with T at 100 K, 200 K, 300 K, 400 K, 500 K are shown in Figure 7 (e) and (f), respectively. $\alpha$ increases rapidly with T at low temperature and decreases non-linearly with P at low pressure. In addition, $\alpha$ gradually approaches a linear increase to T at high temperature ($T\geq 400 K$) and a linear decrease to P at high pressure ($T\geq 15 GPa$). So, the effects of T and P on $\alpha$ are opposite as well.

So in conclusion, we have found that the effects of T and P on $C_V$, $\theta_D$ and $\alpha$ are opposite.

 \subsection{Structural Stability and well as Coulomb Effect}
In this section, we study the stability of the cubic structure under tetragonal distortion at first. The energy of the tetragonal phase as a function of c/a is shown in Figure 8. The ratio c/a=1 represents the cubic geometry. As can be seen, cubic structure has minimum energy in a wide range of the ratio c/a. So the cubic structure of Ba$_2$PrVO$_6$ is more stable than its tetragonal phase. 
 
Finally let us study Coulomb effect on the properties of Ba$_2$PrVO$_6$. The total and partial density of states for some values of U ($U_V$ = 3 and 4 eV; $U_{Pr}$=8 and 9 eV) are shown in Figure 9. One can see that the total and partial density of states are consistent in different values of U. It indicates that the compensated half metallicity can be preserved when the Coulomb interaction is taken into account in this case.

\section{Conclusions}
In the present paper, the electronic structure, band structures, magnetic properties, magnetic evaluation and thermodynamic properties of the double perovskite compound Ba$_2$PrVO$_6$ have been investigated by using first-principle calculations. Within LSDA calculation, Ba$_2$PrVO$_6$ is metallic ferromagnetic. It is found that Ba$_2$PrVO$_6$ is a compensated half metal under LSDA+U calculations with the anti-ferromagnetism of Ba$_2$PrVO$_6$ is originated from the anti-aligned spin states between ($4f_{-2}$) state and hybridized $t_{2g}-\alpha$ state. Through FSM calculations, we find that compensated half metallic state is the ground state. Employing the quasi-harmonic Debye model, we have studied the thermodynamic properties of Ba$_2$PrVO$_6$, such as bulk modulus, heat capacity, Debye temperature and thermal expansion under temperature and pressure. We hope that our work will stimulate the exploration of CHM  in both experiment and theory.

\newpage
\begin{table}[!hbp]
\caption{\label{arttype}Total and partial magnetic moments for Ba$_2$PrVO$_6$ within LSDA and LSDA+U calculations. The bond lengths are also shown in this table.}
\begin{tabular*}{\textwidth}{@{}l*{15}{@{\extracolsep{0pt plus12pt}}l}}
\hline
\hline
	&$M_{tot}$&$M_{Pr}$&$M_V$&$M_{6O}$\\
LSDA&0.74&0.79&0.05&-0.11&\\
LSDA+U&0.00&-0.63&0.47&0.19\\
	\hline
	Bond &Pr-O&V-O\\
	bond-length (\AA)  &1.951&2.210\\
	\hline
	\hline
\end{tabular*}
\end{table}

\newpage
\begin{figure}[htp]

\includegraphics[width=8in]{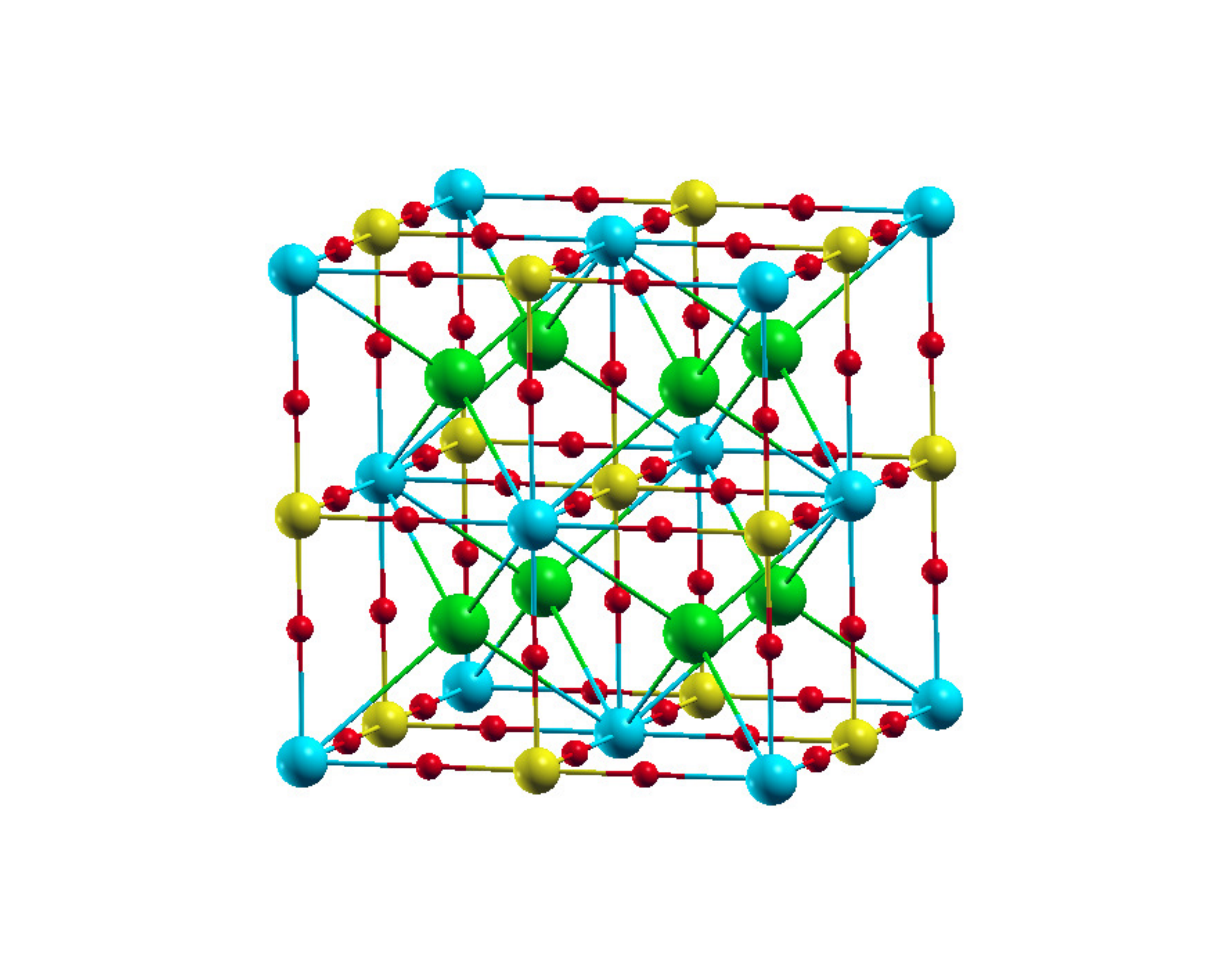}
\caption{Crystal structure for Ba$_2$PrVO$_6$. Green, blue yellow and red spheres represent Ba, Pr, V and O atoms, respectively.}

\end{figure}

\newpage

\begin{figure}[!htb]
\centering
\subfigure[]{
\begin{minipage}[c]{0.45\textwidth}
\includegraphics[width=1.2\textwidth,angle=-90]{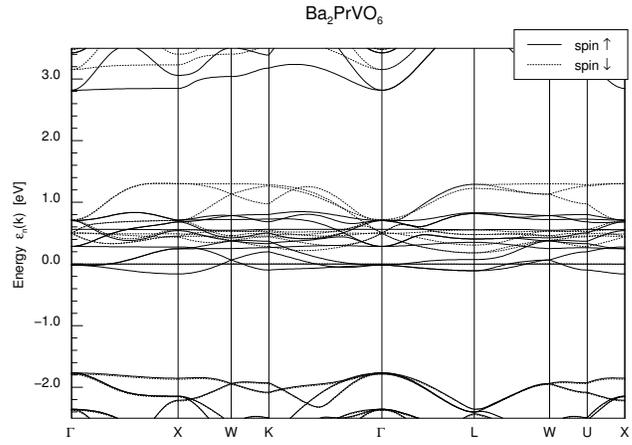} \\
\end{minipage}
}

\subfigure[]{
\begin{minipage}[c]{0.45\textwidth}
\includegraphics[width=1.2\textwidth,angle=-90]{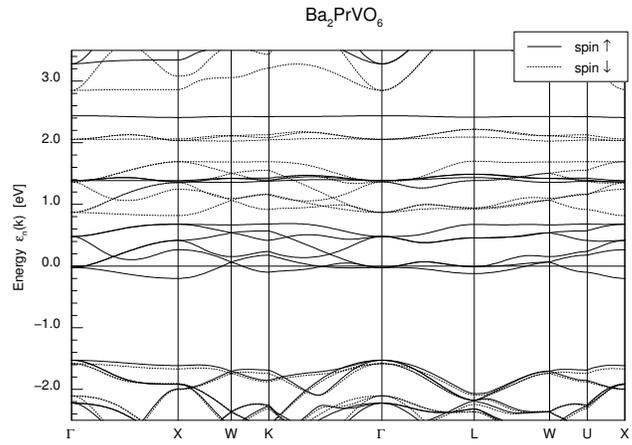} \\
\end{minipage}
}

\caption{Band structures of Ba$_2$PrVO$_6$ within (a)LSDA and (b) LSDA+U.}
\end{figure}
\newpage

\begin{figure}[!htb]
\centering
\subfigure[]{
\begin{minipage}[c]{0.45\textwidth}
\includegraphics[width=1.2\textwidth,angle=0]{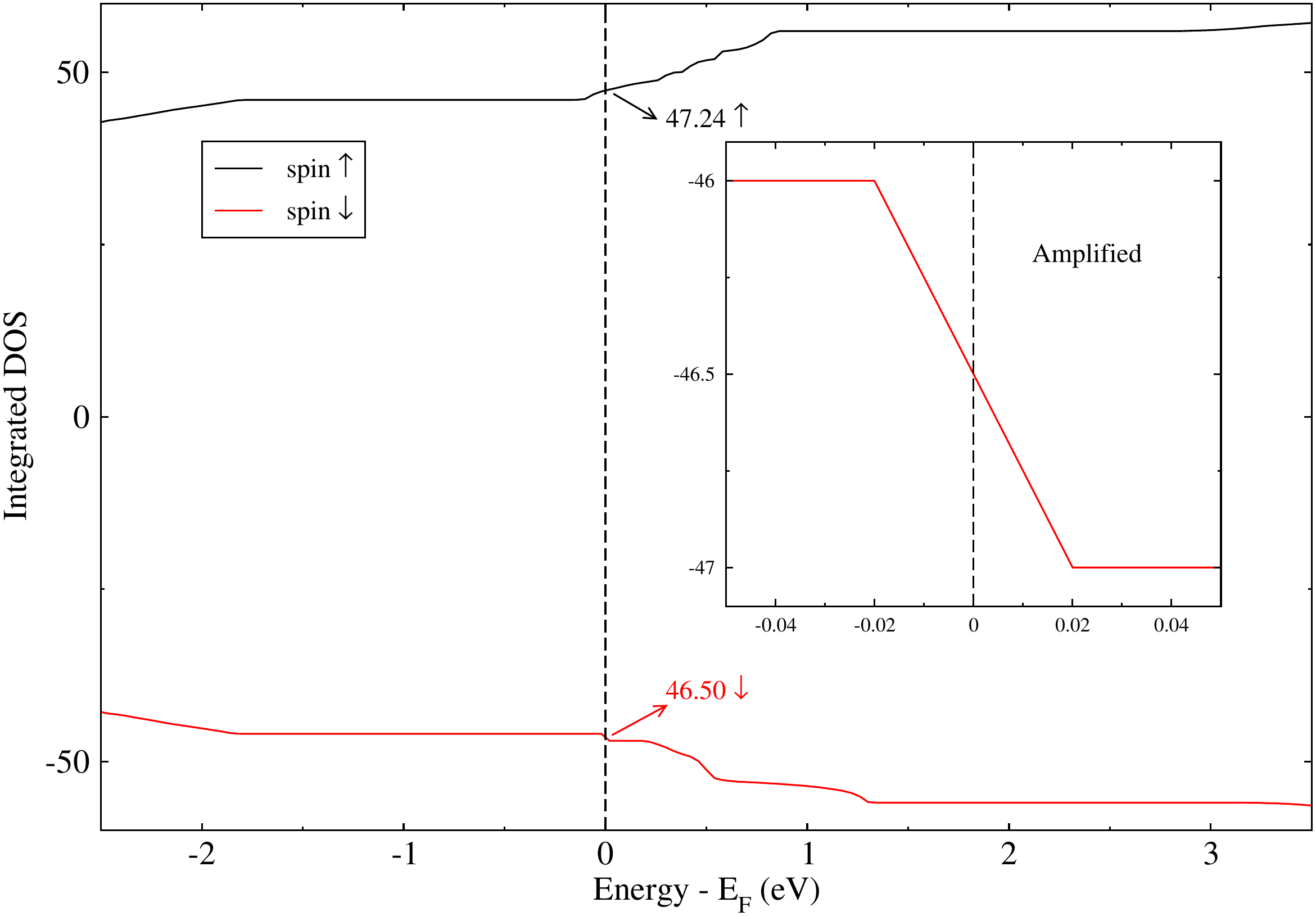} \\
\end{minipage}
}

\subfigure[]{
\begin{minipage}[c]{0.45\textwidth}
\includegraphics[width=1.2\textwidth,angle=0]{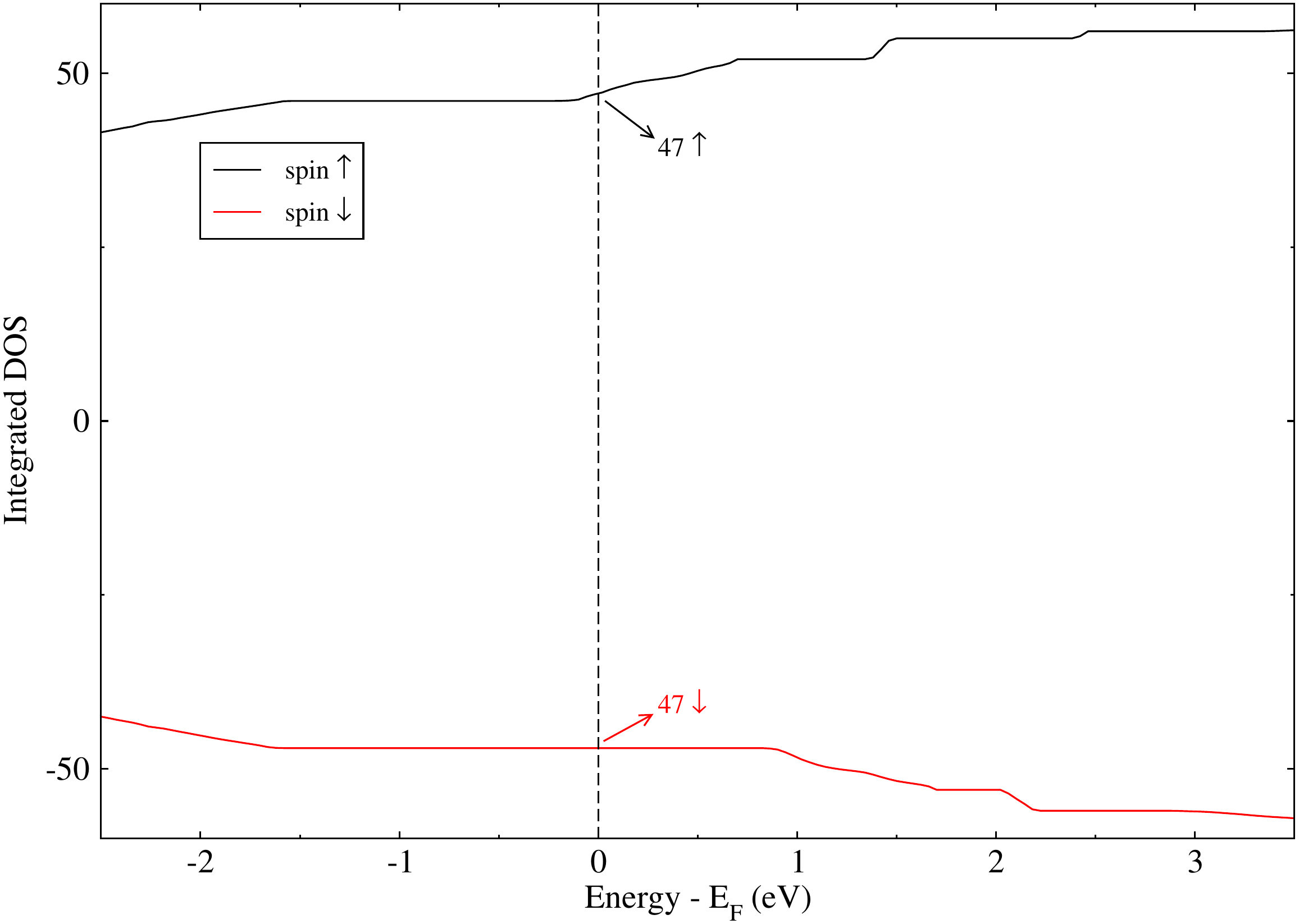} \\
\end{minipage}
}

\caption{(a) Integrated DOS plot of Ba$_2$PrVO$_6$ within LSDA. The amplified spin down integrated DOS plot near the Fermi level is inserted at right hand. (b)
Integrated DOS plot of $Ba_2PrVO_6$ within LSDA+U.}
\end{figure}

\newpage

\begin{figure}[htp]

\includegraphics[width=6.8in]{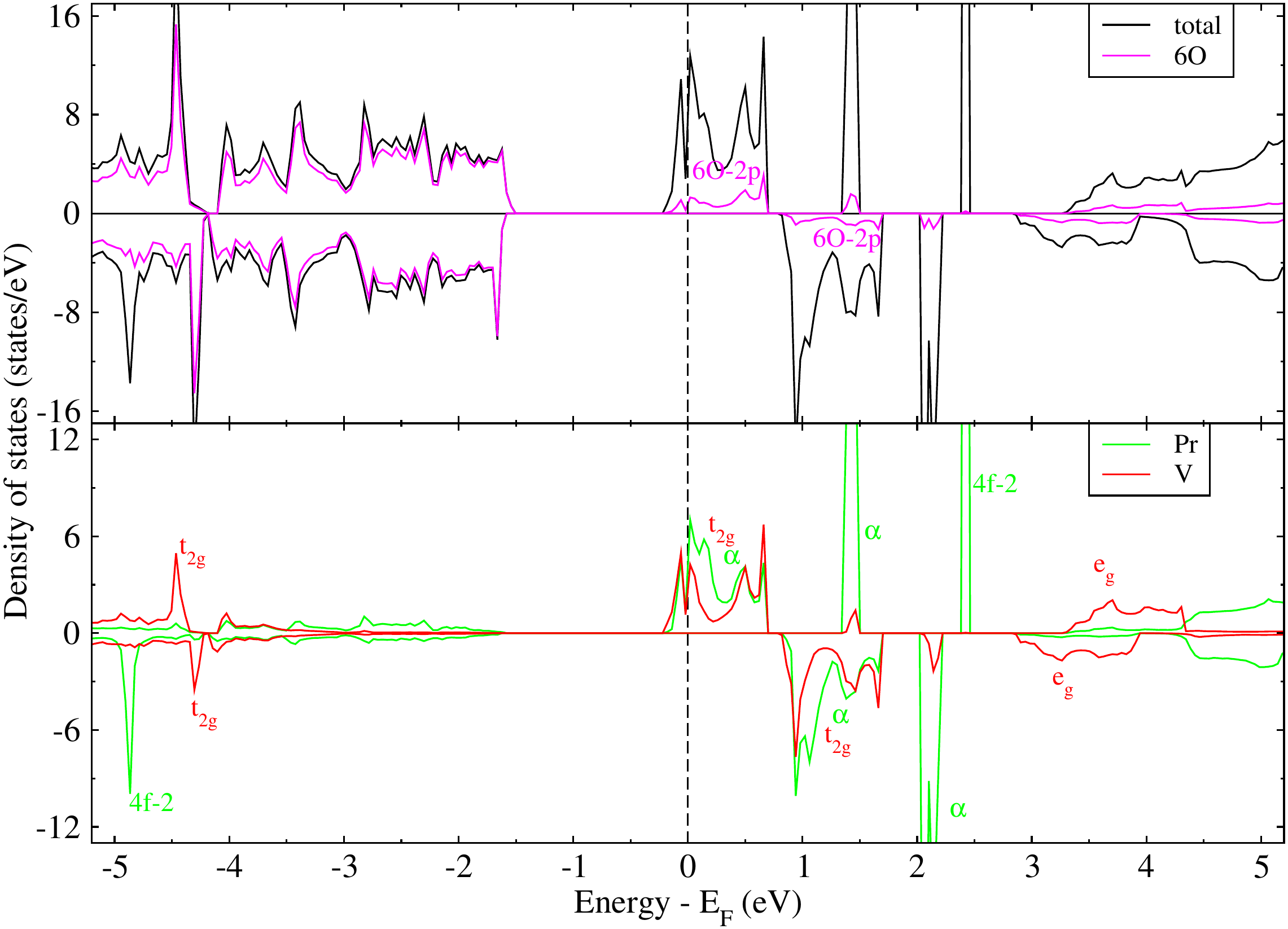}
\caption{Total and partial density of states for Ba$_2$PrVO$_6$ within LSDA+U. The vertical dashed line denotes Fermi level, which is set to zero.}

\end{figure}

\newpage
\begin{figure}[htp]

\includegraphics[width=6.8in]{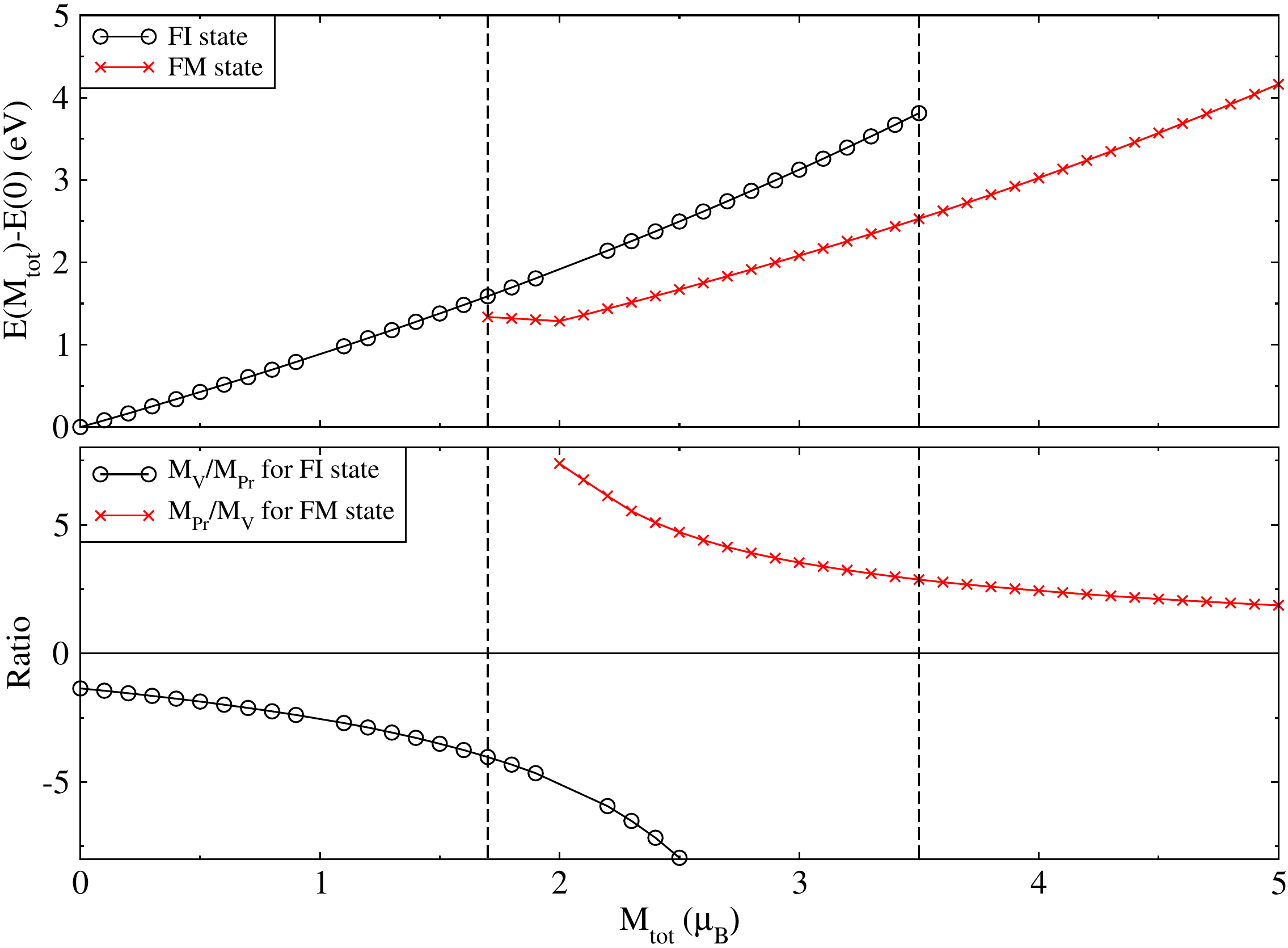}
\caption{Fixed spin moment calculations within LSDA+U for Ba$_2$PrVO$_6$. Top: energy vs total magnetic moment ($M_{tot}$) plot. The zero energy E(0) denotes the previous calculated compensated Ba$_2$PrVO$_6$. Bottom: the ratio plots between the magnetic moment of V ($M_V$) and that of Pr ($M_V$). For both up and bottom plots, the black circle and red multiplicative symbol plots represent the ferrimagnetic (FI) and ferromagnetic (FM) states, respectively. The FI and FM states coexist in the region between the broken lines. For bottom plot, the black circle and red multiplicative symbol plots represent the ratios of $\frac{M_V}{M_{Pr}}$ for FI state and $\frac{M_{Pr}}{M_{V}}$ for FM state, respectively.}

\end{figure}

\newpage
\begin{figure}[htp]

\includegraphics[width=6.8in]{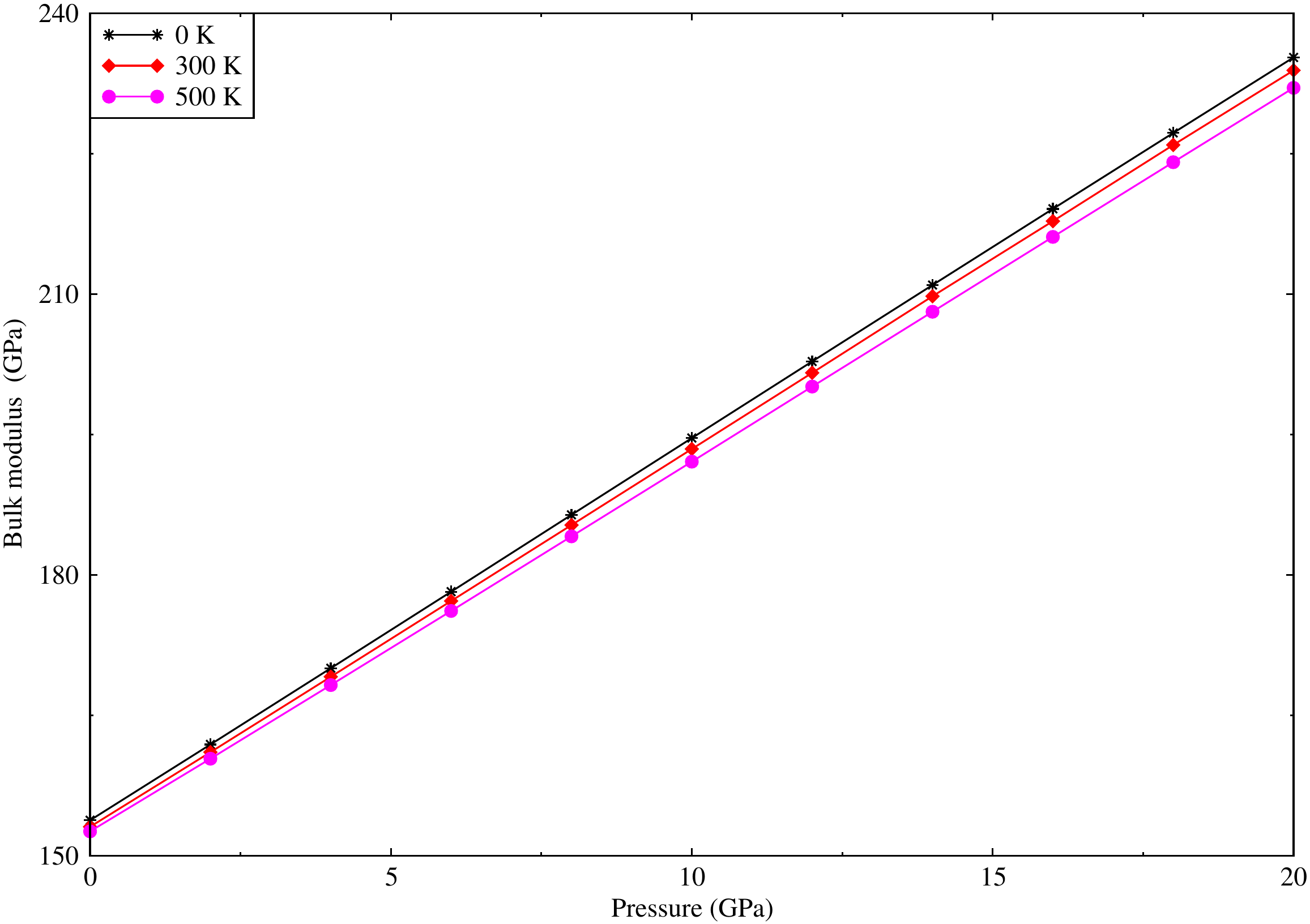}
\caption{Bulk modulus of Ba$_2$PrVO$_6$ as a function of pressure at 0 K, 300 K and 500 K.}

\end{figure}

\newpage

\begin{figure}[!htb]
\centering
\subfigure[]{
\begin{minipage}[c]{0.45\textwidth}
\includegraphics[width=1\textwidth,angle=0]{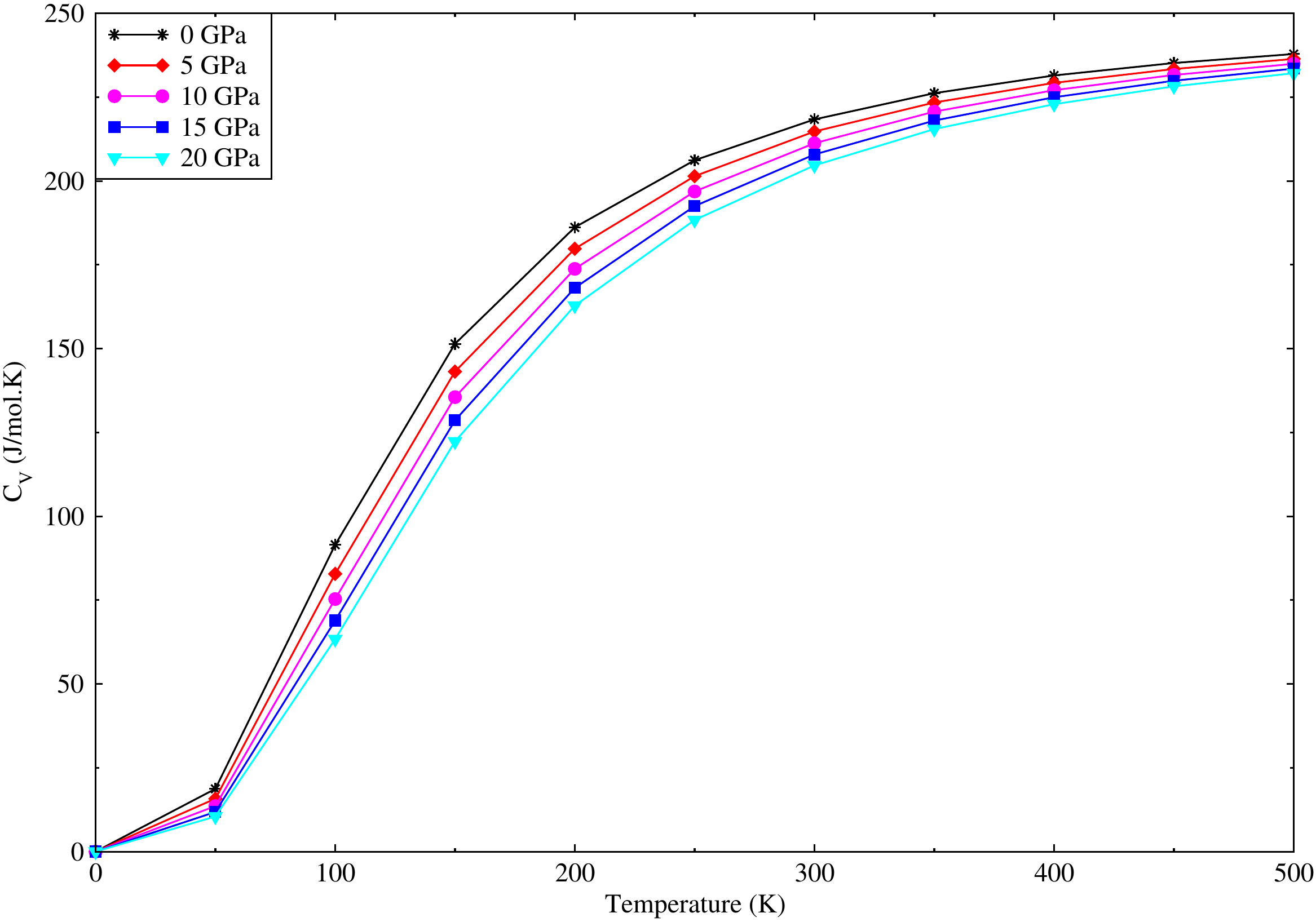} \\
\end{minipage}
}
\subfigure[]{
\begin{minipage}[c]{0.45\textwidth}
\includegraphics[width=1\textwidth,angle=0]{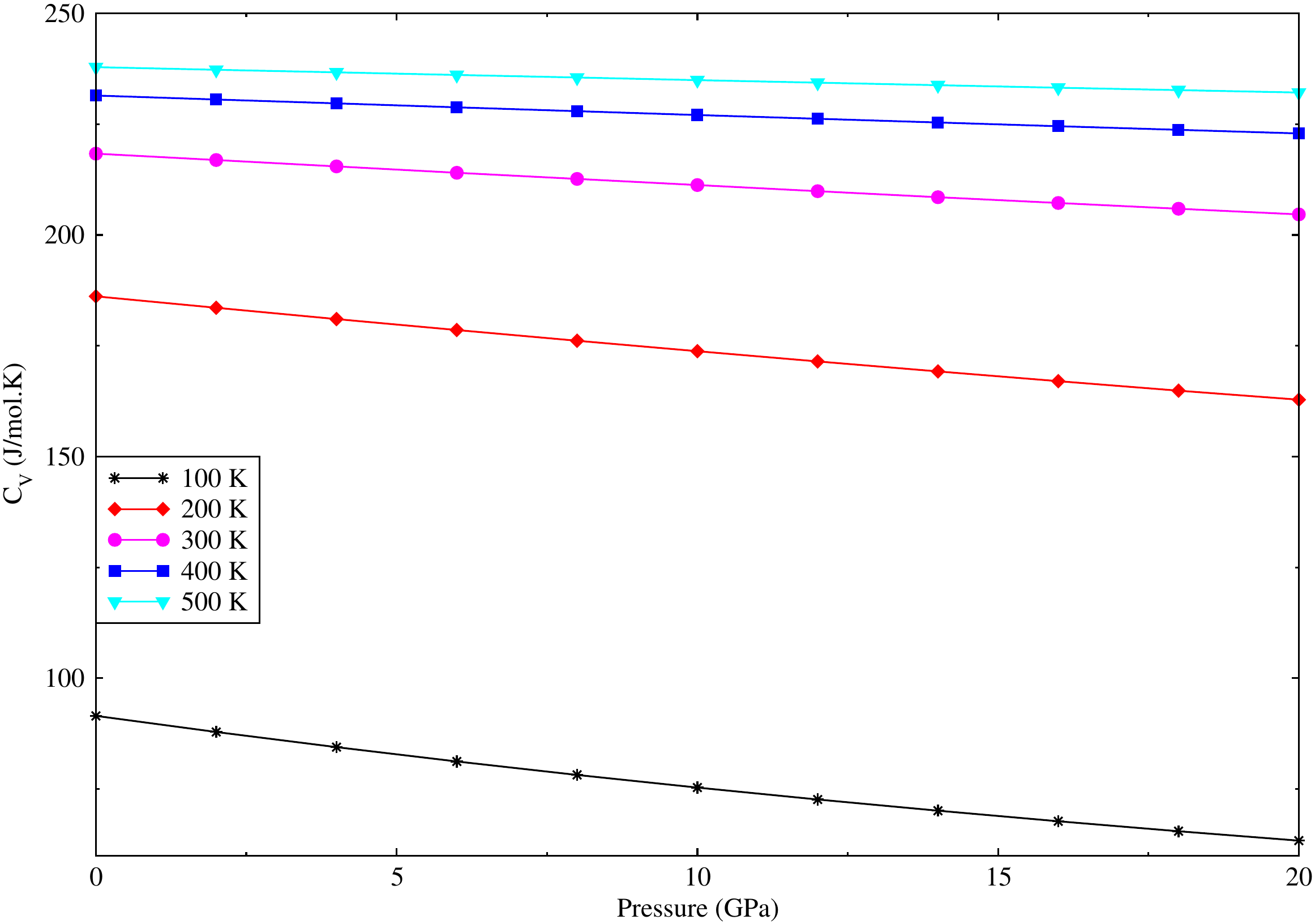} \\
\end{minipage}
}

\subfigure[]{
\begin{minipage}[c]{0.45\textwidth}
\includegraphics[width=1\textwidth,angle=0]{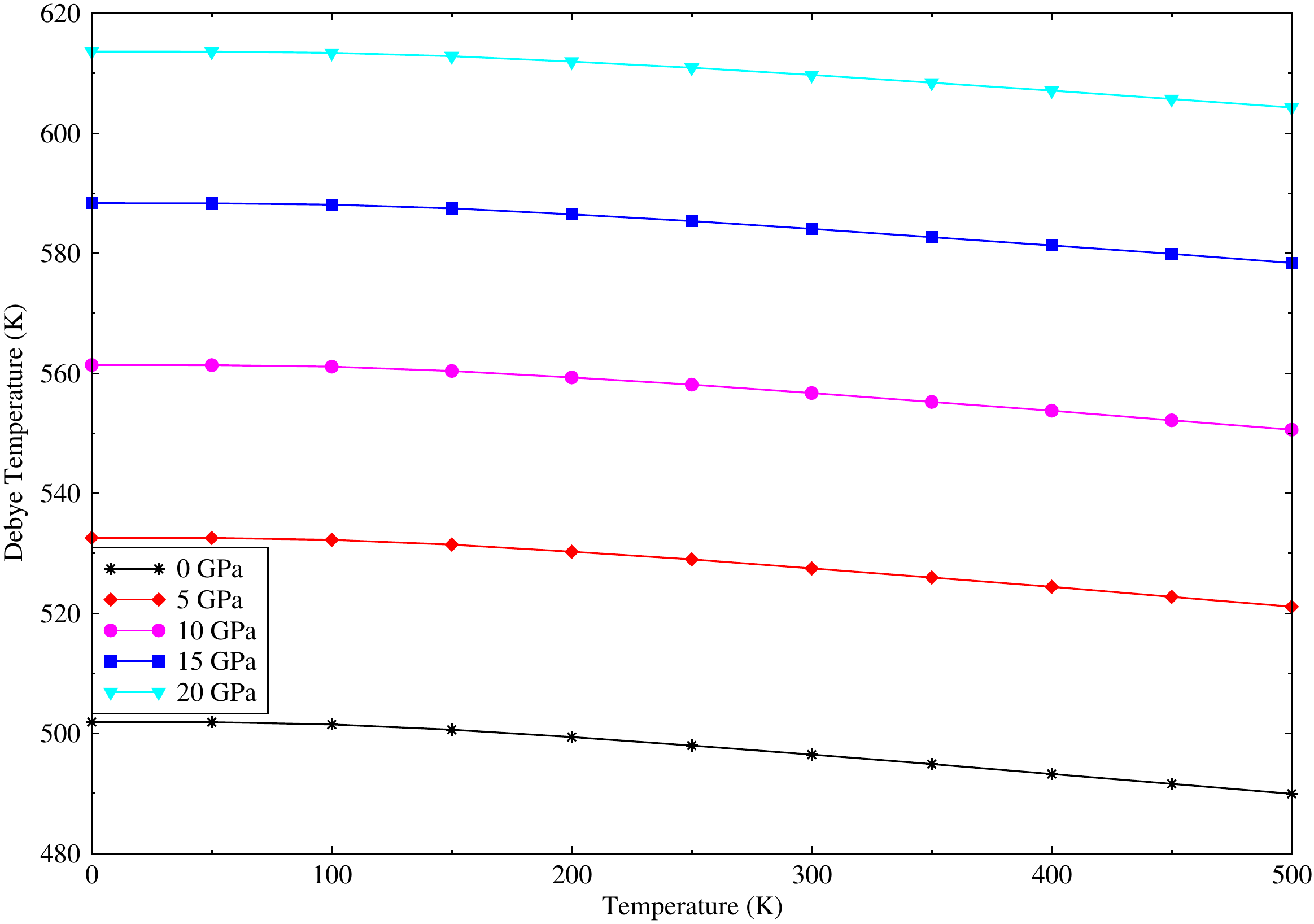} \\
\end{minipage}
}
\subfigure[]{
\begin{minipage}[c]{0.45\textwidth}
\includegraphics[width=1\textwidth,angle=0]{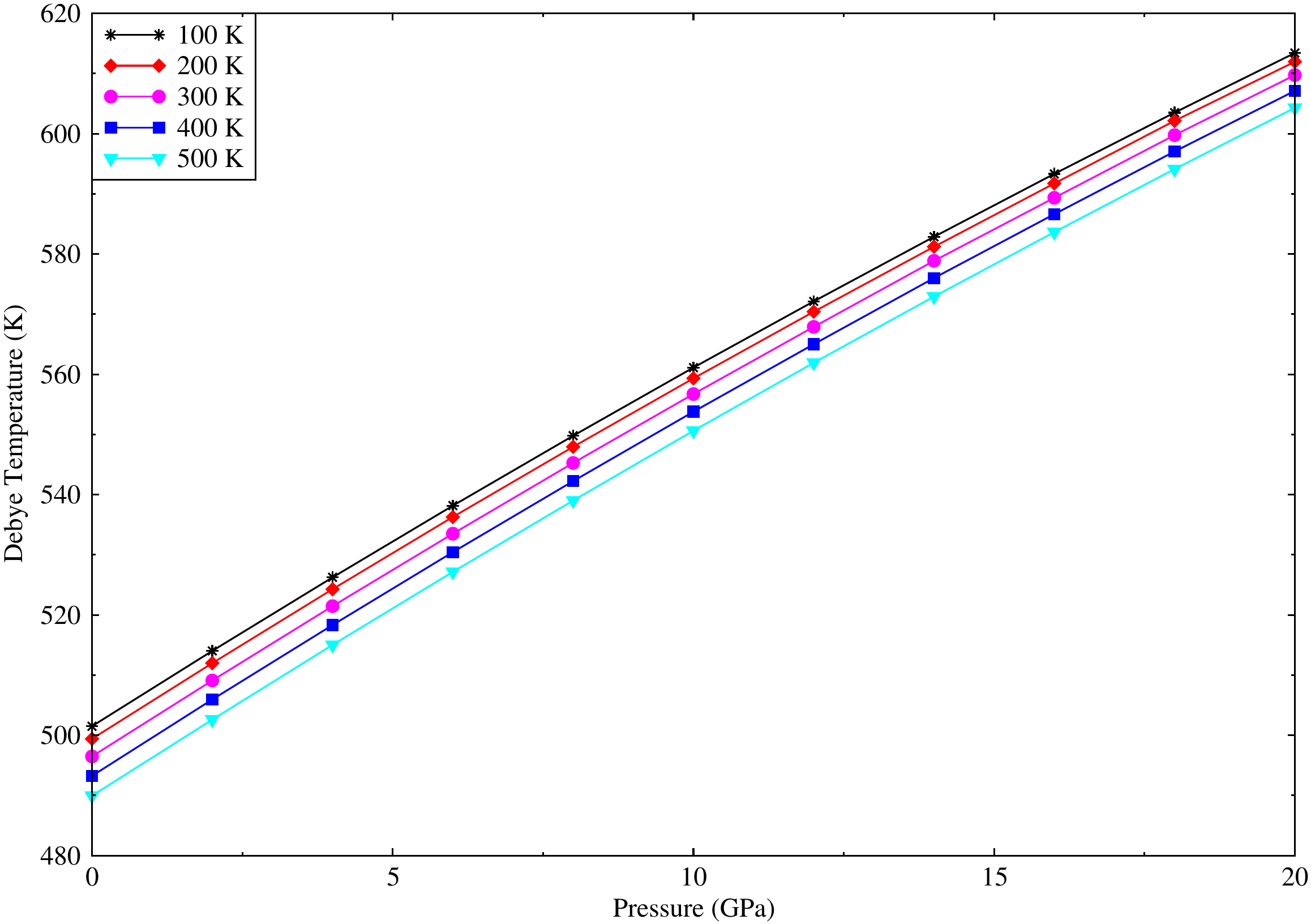} \\
\end{minipage}
}

\subfigure[]{
\begin{minipage}[c]{0.45\textwidth}
\includegraphics[width=1\textwidth,angle=0]{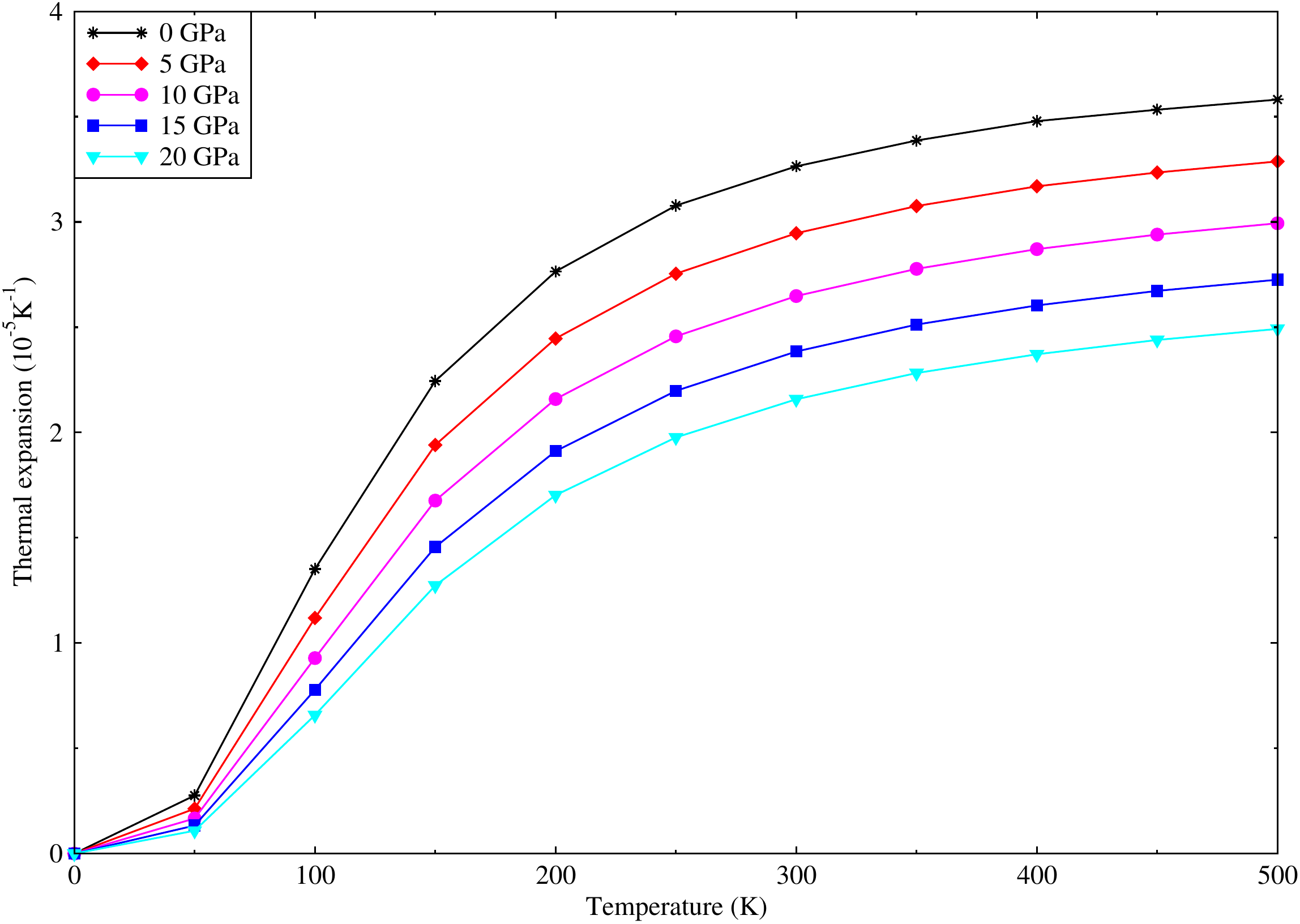} \\
\end{minipage}
}
\subfigure[]{
\begin{minipage}[c]{0.45\textwidth}
\includegraphics[width=1\textwidth,angle=0]{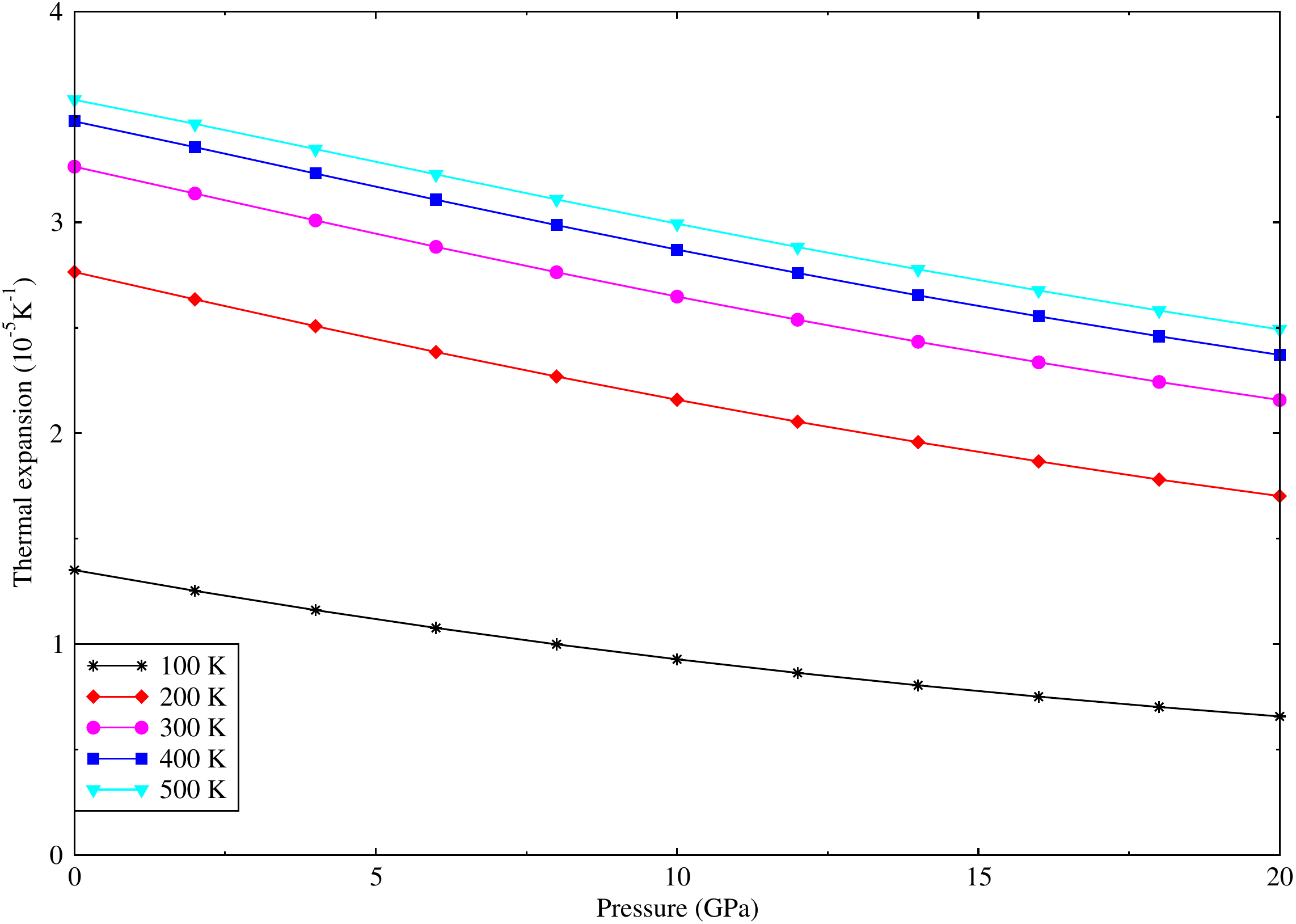} \\
\end{minipage}
}

\caption{ Heat capacity ($C_V$) (a), Debye temperature (c) and thermal expansion (e) as a function of temperature at 0 GPa, 5 GPa, 10 GPa, 15 GPa and 20 GPa. Heat capacity ($C_V$) (b), Debye temperature (d) and thermal expansion (f) as a function of pressure at 100 K, 200 K, 300 K, 400 K and 500 K.}
\end{figure}

\newpage
\begin{figure}[htp]

\includegraphics[width=6.8in]{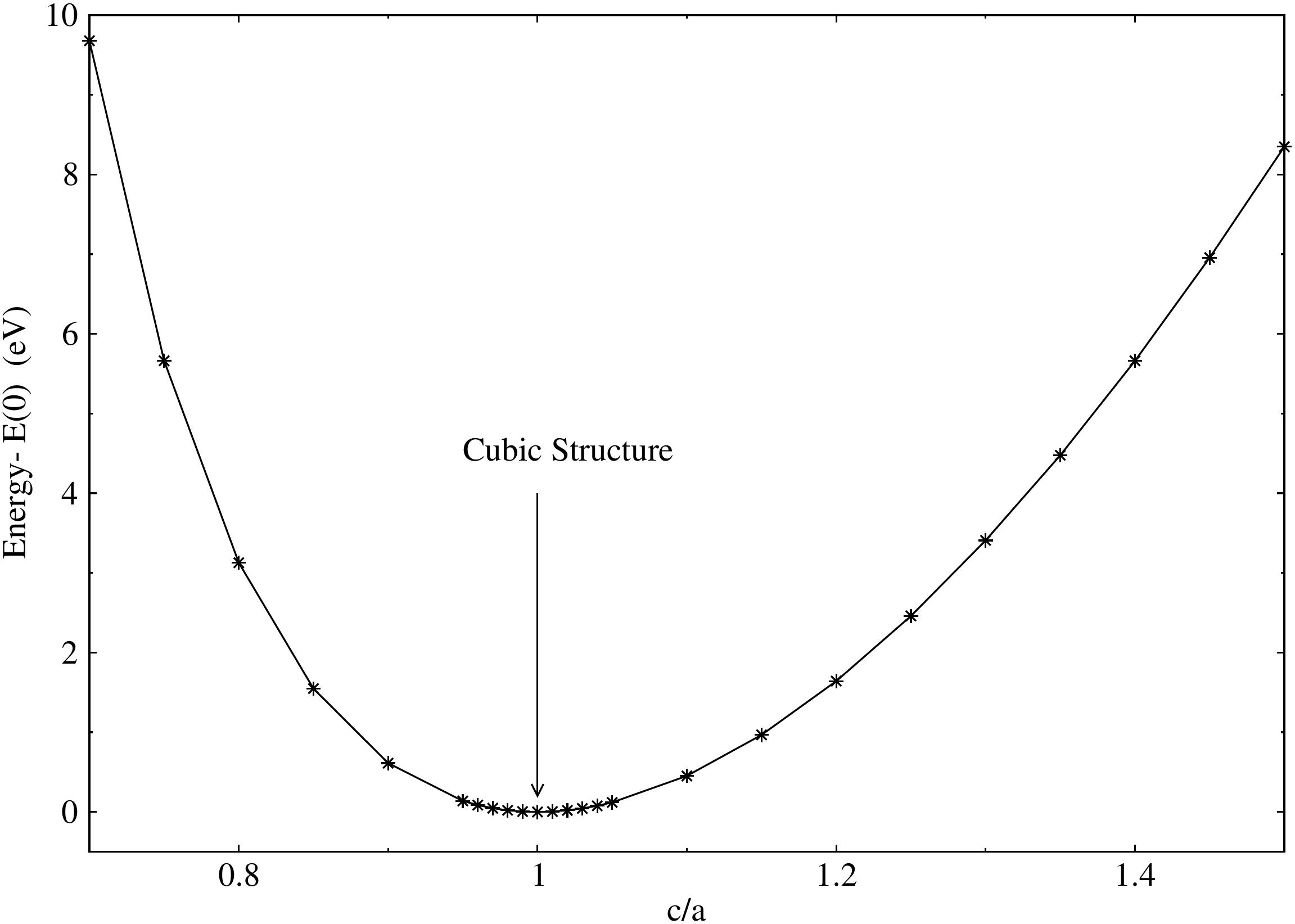}
\caption{Total energy under tetragonal distortion as a function of c/a ratio at constant volume. E(0) is the energy of cubic structure.}

\end{figure}

\newpage

\begin{figure}[!htb]
\centering
\subfigure[]{
\begin{minipage}[c]{0.45\textwidth}
\includegraphics[width=1\textwidth,angle=0]{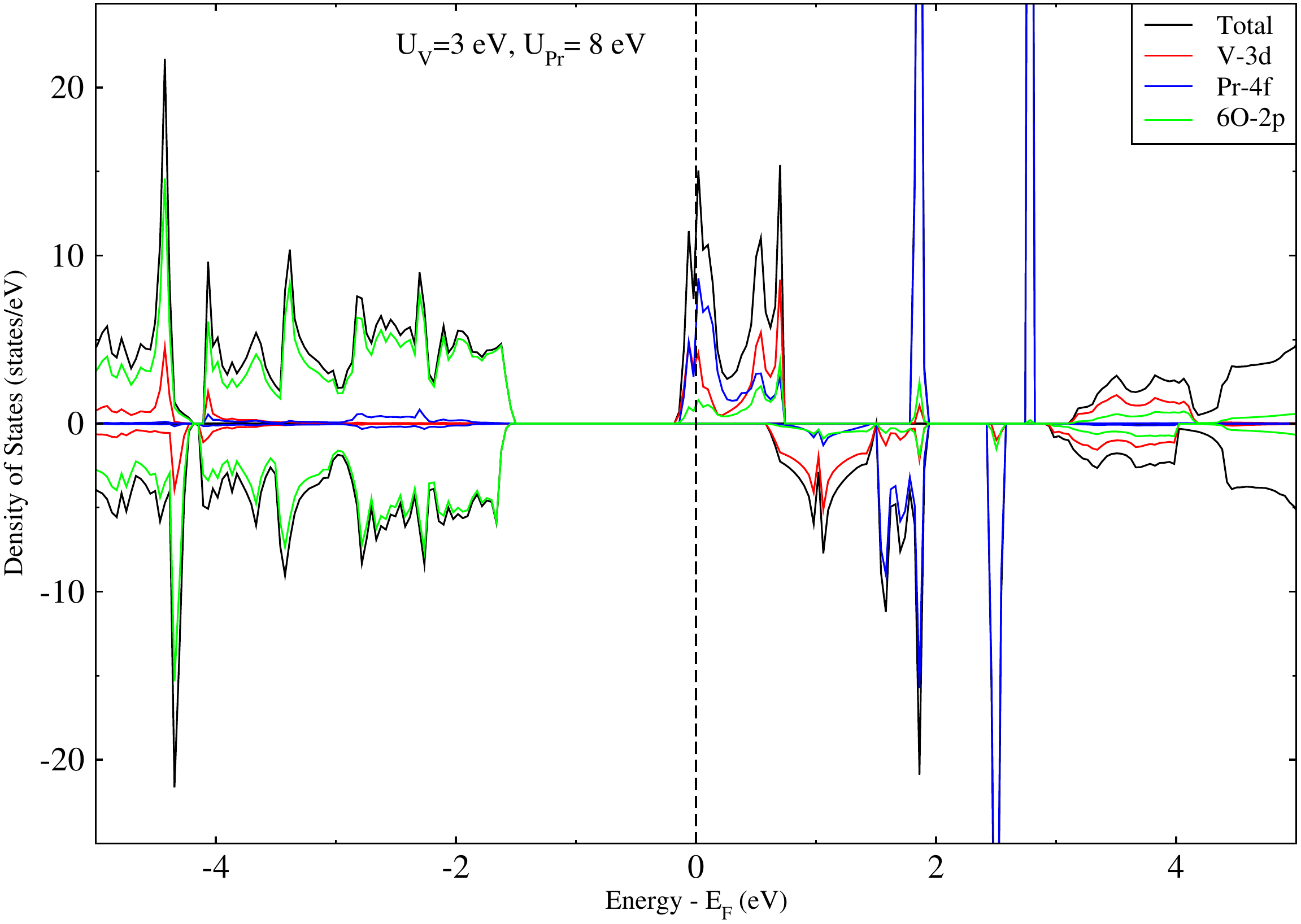} \\
\end{minipage}
}
\subfigure[]{
\begin{minipage}[c]{0.45\textwidth}
\includegraphics[width=1\textwidth,angle=0]{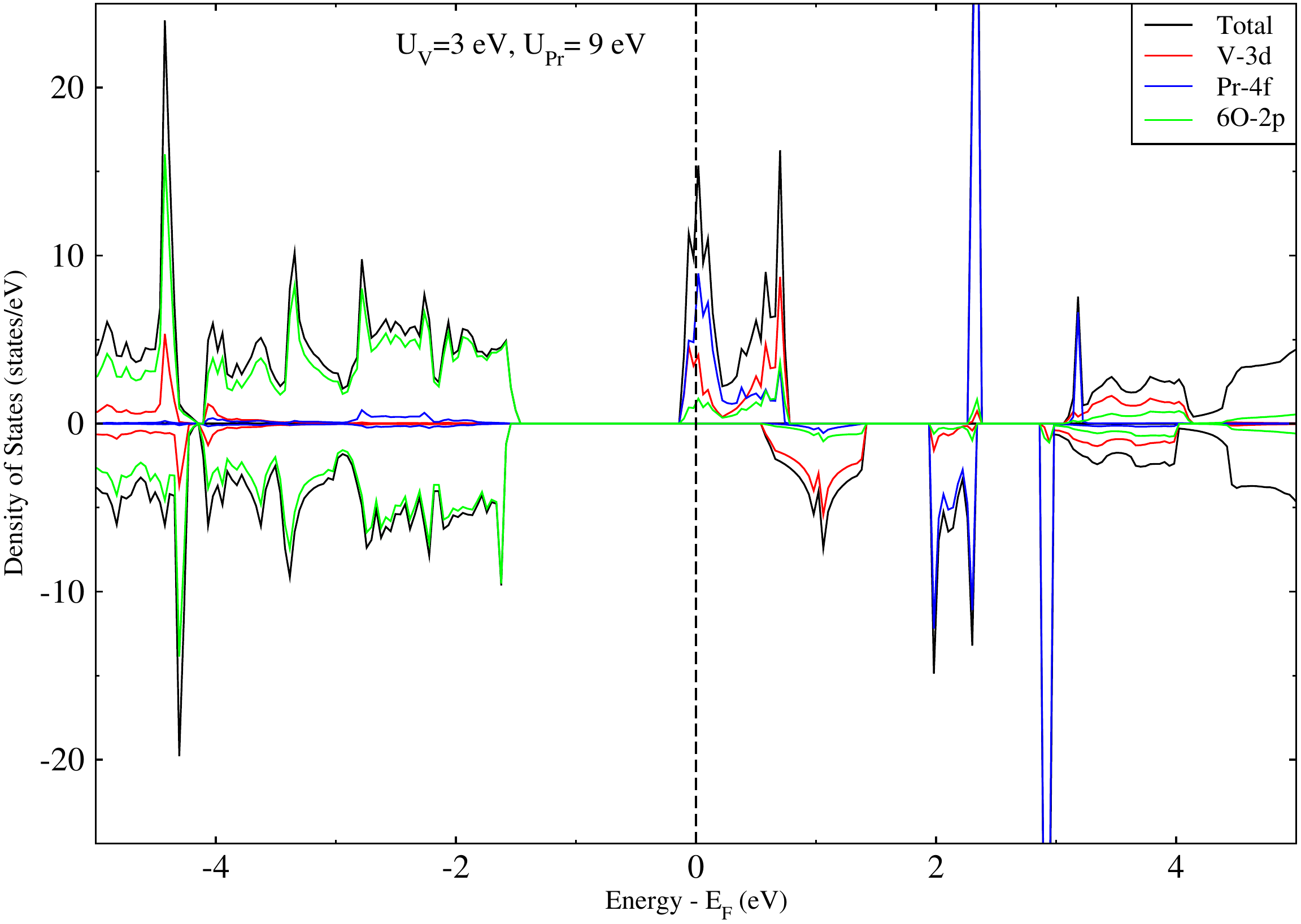} \\
\end{minipage}
}

\subfigure[]{
\begin{minipage}[c]{0.45\textwidth}
\includegraphics[width=1\textwidth,angle=0]{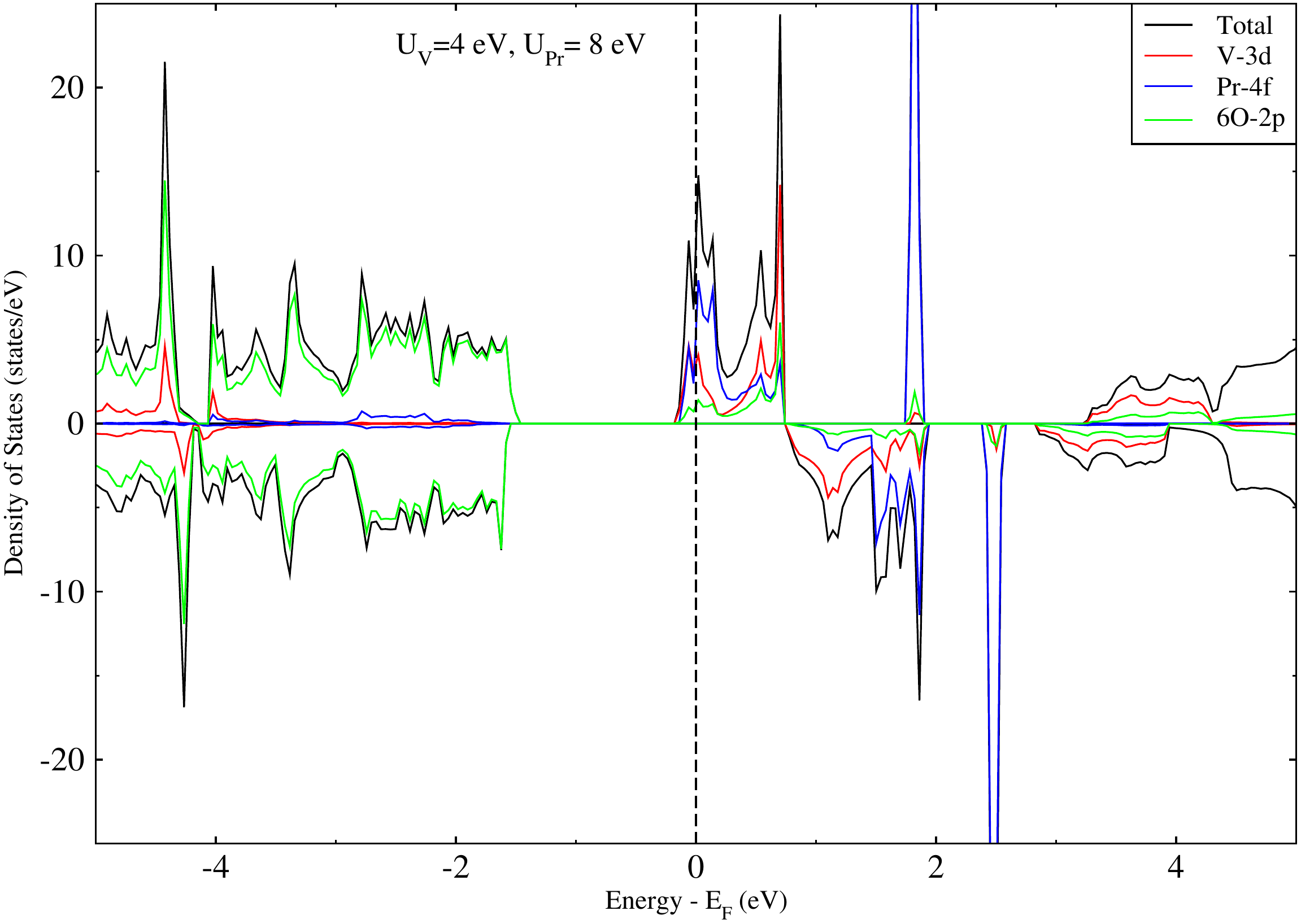} \\
\end{minipage}
}
\subfigure[]{
\begin{minipage}[c]{0.45\textwidth}
\includegraphics[width=1\textwidth,angle=0]{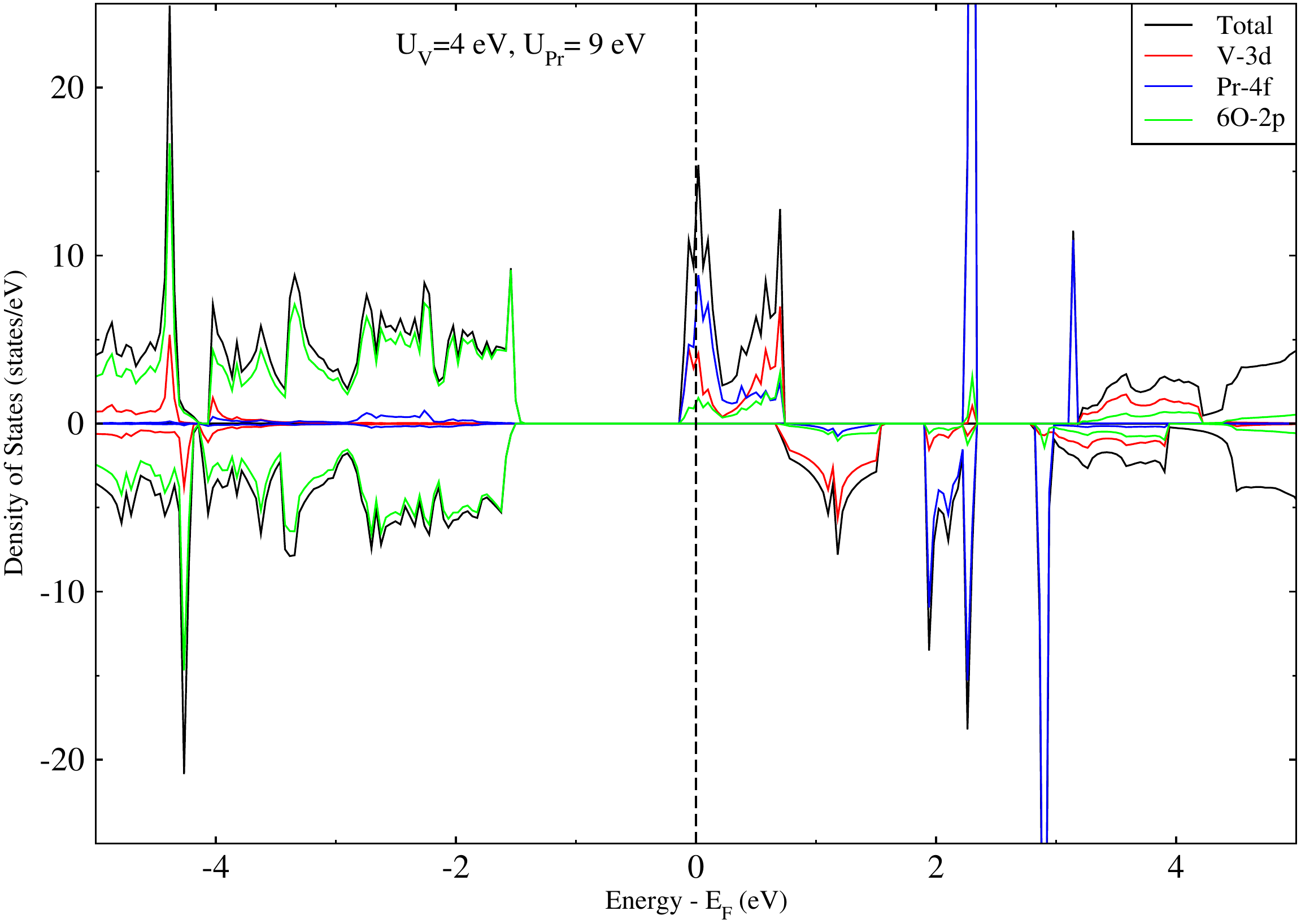} \\
\end{minipage}
}

\caption{ Total and partial density of states for (a) $U_V = 3$ eV,  $U_{Pr}$ = 8 eV; (b) $U_V$ = 3 eV,  $U_{Pr}$ = 9 eV; (c) $U_V$ = 4 eV,  $U_{Pr}$ = 8 eV; (d) $U_V$ = 4 eV,  $U_{Pr}$ = 9 eV.}
\end{figure}


\begin{thebibliography}{10}
\expandafter\ifx\csname url\endcsname\relax
  \def\url#1{\texttt{#1}}\fi
\expandafter\ifx\csname urlprefix\endcsname\relax\def\urlprefix{URL }\fi
\expandafter\ifx\csname href\endcsname\relax
  \def\href#1#2{#2} \def\path#1{#1}\fi

\bibitem{1}
M. I. Katsnelson, V. Yu. Irkhin, L. Chioncel, A. I. Lichtenstein, and R. A. de Groot,
\href{http://link.aps.org/doi/10.1103/RevModPhys.80.315}{Half-metallic
  ferromagnets: From band structure to many-body effects} 80~(2).
\newblock \href {http://dx.doi.org/10.1103/RevModPhys.80.315}
  {\path{doi:10.1103/RevModPhys.80.315}}.
\newline\urlprefix\url{http://link.aps.org/doi/10.1103/RevModPhys.80.315}

\bibitem{2}
G.~Schmidt, D.~Ferrand, L.~W. Molenkamp, A.~T. Filip, B.~J. van Wees,
  \href{http://link.aps.org/doi/10.1103/PhysRevB.62.R4790}{Fundamental obstacle
  for electrical spin injection from a ferromagnetic metal into a diffusive
  semiconductor}, Physical Review B 62~(8) (2000) R4790--R4793.
\newblock \href {http://dx.doi.org/10.1103/PhysRevB.62.R4790}
  {\path{doi:10.1103/PhysRevB.62.R4790}}.
\newline\urlprefix\url{http://link.aps.org/doi/10.1103/PhysRevB.62.R4790}

\bibitem{3}
R.~A. de~Groot, F.~M. Mueller, P.~G.~v. Engen, K.~H.~J. Buschow,
  \href{http://link.aps.org/doi/10.1103/PhysRevLett.50.2024}{New Class of
  Materials: Half-Metallic Ferromagnets}, Physical Review Letters
  50~(25) (1983) 2024--2027.
\newblock \href {http://dx.doi.org/10.1103/PhysRevLett.50.2024}
  {\path{doi:10.1103/PhysRevLett.50.2024}}.
\newline\urlprefix\url{http://link.aps.org/doi/10.1103/PhysRevLett.50.2024}

\bibitem{4}
F.~Casper, H.~C. Kandpal, G.~H. Fecher, C.~Felser,
  \href{http://stacks.iop.org/0022-3727/40/i=10/a=002?key=crossref.451b8b984fc845faaca89eb9cb51e1d6}{Electronic and magnetic properties of GdPdSb}, Journal of Physics D: Applied Physics
  40~(10) (2007) 3024--3029.
\newblock \href {http://dx.doi.org/10.1088/0022-3727/40/10/002}
  {\path{doi:10.1088/0022-3727/40/10/002}}.
\newline\urlprefix\url{http://stacks.iop.org/0022-3727/40/i=10/a=002?key=crossref.451b8b984fc845faaca89eb9cb51e1d6}

\bibitem{5}
I.~Galanakis, K.~Özdoğan, E.~Şaşıog̃lu,
  \href{http://stacks.iop.org/0953-8984/26/i=8/a=086003?key=crossref.d71e76d690676543fd6db7d726de38d5}{High-T$_C$ fully compensated ferrimagnetic semiconductors as spin-filter materials: the case of CrVXAl (X = Ti, Zr, Hf) Heusler compounds}, Journal of Physics: Condensed Matter 26~(8) (2014)
  086003.
\newblock \href {http://dx.doi.org/10.1088/0953-8984/26/8/086003}
  {\path{doi:10.1088/0953-8984/26/8/086003}}.
\newline\urlprefix\url{http://stacks.iop.org/0953-8984/26/i=8/a=086003?key=crossref.d71e76d690676543fd6db7d726de38d5}

\bibitem{6}
Jean-Baptiste Moussy,
\href{http://iopscience.iop.org/article/10.1088/0022-3727/46/14/143001/meta}{From
  epitaxial growth of ferrite thin films to spin-polarized tunnelling }, Journal of Physics D: Applied Physics 46~(14) (2013).
\newline\urlprefix\url{http://iopscience.iop.org/article/10.1088/0022-3727/46/14/143001/meta}

\bibitem{7}
E.~N. Voloshina, A.~Generalov, M.~Weser, S.~Böttcher, K.~Horn, Y.~S. Dedkov,
  \href{http://stacks.iop.org/1367-2630/13/i=11/a=113028?key=crossref.448df8281f40a9c9ba2cec5dee576e46}{Structural
  and electronic properties of the graphene/{Al}/{Ni}(111) intercalation
  system}, New Journal of Physics 13~(11) (2011) 113028.
\newblock \href {http://dx.doi.org/10.1088/1367-2630/13/11/113028}
  {\path{doi:10.1088/1367-2630/13/11/113028}}.
\newline\urlprefix\url{http://stacks.iop.org/1367-2630/13/i=11/a=113028?key=crossref.448df8281f40a9c9ba2cec5dee576e46}

\bibitem{8}
H.~Kurt, K.~Rode, P.~Stamenov, M.~Venkatesan, Y.-C. Lau, E.~Fonda, J.~M.~D.
  Coey, \href{http://link.aps.org/doi/10.1103/PhysRevLett.112.027201}{Cubic $Mn_2Ga$
  Thin Films: Crossing the Spin Gap with Ruthenium}, Physical
  Review Letters 112~(2) (2014) 027201.
\newblock \href {http://dx.doi.org/10.1103/PhysRevLett.112.027201}
  {\path{doi:10.1103/PhysRevLett.112.027201}}.
\newline\urlprefix\url{http://link.aps.org/doi/10.1103/PhysRevLett.112.027201}

\bibitem{9}
G.~M. Müller, J.~Walowski, M.~Djordjevic, G.-X. Miao, A.~Gupta, A.~V. Ramos,
  K.~Gehrke, V.~Moshnyaga, K.~Samwer, J.~Schmalhorst, A.~Thomas, A.~Hütten,
  G.~Reiss, J.~S. Moodera, M.~Münzenberg,
  \href{http://www.nature.com/nmat/journal/v8/n1/abs/nmat2341.html}{Spin
  polarization in half-metals probed by femtosecond spin excitation}, Nature
  Materials 8~(1) (2009) 56--61.
\newblock \href {http://dx.doi.org/10.1038/nmat2341}
  {\path{doi:10.1038/nmat2341}}.
\newline\urlprefix\url{http://www.nature.com/nmat/journal/v8/n1/abs/nmat2341.html}

\bibitem{10}
H.~van Leuken, R.~A. de~Groot,
  \href{http://link.aps.org/doi/10.1103/PhysRevLett.74.1171}{Half-{Metallic}
  {Antiferromagnets}}, Physical Review Letters 74~(7) (1995) 1171--1173.
\newblock \href {http://dx.doi.org/10.1103/PhysRevLett.74.1171}
  {\path{doi:10.1103/PhysRevLett.74.1171}}.
\newline\urlprefix\url{http://link.aps.org/doi/10.1103/PhysRevLett.74.1171}

\bibitem{11}
S.~Wurmehl, H.~C. Kandpal, G.~H. Fecher, C.~Felser,
  \href{http://stacks.iop.org/0953-8984/18/i=27/a=001?key=crossref.a1520d99c63b42b546ceee50733ebca1}{Valence
  electron rules for prediction of half-metallic compensated-ferrimagnetic
  behaviour of {Heusler} compounds with complete spin polarization}, Journal of
  Physics: Condensed Matter 18~(27) (2006) 6171--6181.
\newblock \href {http://dx.doi.org/10.1088/0953-8984/18/27/001}
  {\path{doi:10.1088/0953-8984/18/27/001}}.
\newline\urlprefix\url{http://stacks.iop.org/0953-8984/18/i=27/a=001?key=crossref.a1520d99c63b42b546ceee50733ebca1}

\bibitem{12}
W.~E. Pickett,
  \href{http://link.aps.org/doi/10.1103/PhysRevLett.77.3185}{Single {Spin}
  {Superconductivity}}, Physical Review Letters 77~(15) (1996) 3185--3188.
\newblock \href {http://dx.doi.org/10.1103/PhysRevLett.77.3185}
  {\path{doi:10.1103/PhysRevLett.77.3185}}.
\newline\urlprefix\url{http://link.aps.org/doi/10.1103/PhysRevLett.77.3185}

\bibitem{13}
K.-W. Lee, K.-H. Ahn,
  \href{http://link.aps.org/doi/10.1103/PhysRevB.85.224404}{Evaluation of
  half-metallic antiferromagnetism in $A_2CrFeO_6$ ($A = La, Sr$)}, Physical Review B 85~(22) (2012) 224404.
\newblock \href {http://dx.doi.org/10.1103/PhysRevB.85.224404}
  {\path{doi:10.1103/PhysRevB.85.224404}}.
\newline\urlprefix\url{http://link.aps.org/doi/10.1103/PhysRevB.85.224404}

\bibitem{14}
M.~S. Park, B.~I. Min,
  \href{http://link.aps.org/doi/10.1103/PhysRevB.71.052405}{Electronic
  structures and magnetic properties of $LaAVMoO_6$ ($A=Ca,Sr,Ba$): {Investigation} of possible half-metallic antiferromagnets}, Physical Review
  B 71~(5) (2005) 052405.
\newblock \href {http://dx.doi.org/10.1103/PhysRevB.71.052405}
  {\path{doi:10.1103/PhysRevB.71.052405}}.
\newline\urlprefix\url{http://link.aps.org/doi/10.1103/PhysRevB.71.052405}

\bibitem{15-a}
A.~K. Nayak, M.~Nicklas, S.~Chadov, P.~Khuntia, C.~Shekhar, A.~Kalache,
  M.~Baenitz, Y.~Skourski, V.~K. Guduru, A.~Puri, U.~Zeitler, J.~M.~D. Coey,
  C.~Felser,
  \href{http://www.nature.com/nmat/journal/v14/n7/full/nmat4248.html}{Design of
  compensated ferrimagnetic {Heusler} alloys for giant tunable exchange bias},
  Nature Materials 14~(7) (2015) 679--684.
\newblock \href {http://dx.doi.org/10.1038/nmat4248}
  {\path{doi:10.1038/nmat4248}}.
\newline\urlprefix\url{http://www.nature.com/nmat/journal/v14/n7/full/nmat4248.html}


\bibitem{15}
L.~Balcells, J.~Navarro, M.~Bibes, A.~Roig, B.~Martı́nez, J.~Fontcuberta,
  \href{http://scitation.aip.org/content/aip/journal/apl/78/6/10.1063/1.1346624}{Cationic
  ordering control of magnetization in $Sr_2FeMoO_6$ double perovskite}, Applied
  Physics Letters 78~(6) (2001) 781--783.
\newblock \href {http://dx.doi.org/10.1063/1.1346624}
  {\path{doi:10.1063/1.1346624}}.
\newline\urlprefix\url{http://scitation.aip.org/content/aip/journal/apl/78/6/10.1063/1.1346624}

\bibitem{16}
W.~E. Pickett,
  \href{http://link.aps.org/doi/10.1103/PhysRevB.57.10613}{Spin-density-functional-based
  search for half-metallic antiferromagnets}, Physical Review B 57~(17) (1998)
  10613--10619.
\newblock \href {http://dx.doi.org/10.1103/PhysRevB.57.10613}
  {\path{doi:10.1103/PhysRevB.57.10613}}.
\newline\urlprefix\url{http://link.aps.org/doi/10.1103/PhysRevB.57.10613}

\bibitem{17}
M.~Uehara, M.~Yamada, Y.~Kimishima,
  \href{http://www.sciencedirect.com/science/article/pii/S0038109803009700}{Physical
  properties of double perovskite compounds $ALaVMoO_6$ ($A=Ca, Sr, Ba$)—a possible half-metallic antiferromagnetic system}, Solid State
  Communications 129~(6) (2004) 385--388.
\newblock \href {http://dx.doi.org/10.1016/j.ssc.2003.11.003}
  {\path{doi:10.1016/j.ssc.2003.11.003}}.
\newline\urlprefix\url{http://www.sciencedirect.com/science/article/pii/S0038109803009700}

\bibitem{18}
M.~Musa Saad H.-E.,
  \href{http://www.sciencedirect.com/science/article/pii/S0038109812002372}{Promising
  half-metallic ferromagnetism in double perovskites $Ba_2VTO_6$ ($T=Nb$ and $Mo$): $Ab$-initio {LMTO}-{ASA} investigations}, Solid State Communications
  152~(14) (2012) 1230--1233.
\newblock \href {http://dx.doi.org/10.1016/j.ssc.2012.04.031}
  {\path{doi:10.1016/j.ssc.2012.04.031}}.
\newline\urlprefix\url{http://www.sciencedirect.com/science/article/pii/S0038109812002372}

\bibitem{19}
S.~Chakraverty, A.~Ohtomo, D.~Okuyama, M.~Saito, M.~Okude, R.~Kumai, T.~Arima,
  Y.~Tokura, S.~Tsukimoto, Y.~Ikuhara, M.~Kawasaki,
  \href{http://link.aps.org/doi/10.1103/PhysRevB.84.064436}{Ferrimagnetism and
  spontaneous ordering of transition metals in double perovskite
   $La_2CrFeO_6$ films}, Physical Review B 84~(6)
  (2011) 064436.
\newblock \href {http://dx.doi.org/10.1103/PhysRevB.84.064436}
  {\path{doi:10.1103/PhysRevB.84.064436}}.
\newline\urlprefix\url{http://link.aps.org/doi/10.1103/PhysRevB.84.064436}

\bibitem{20}
K.-W. Lee, W.~E. Pickett,
  \href{http://link.aps.org/doi/10.1103/PhysRevB.77.115101}{Half semimetallic
  antiferromagnetism in the $Sr_2CrTO_6$ system ($T=Os, Ru$)},
  Physical Review B 77~(11) (2008) 115101.
\newblock \href {http://dx.doi.org/10.1103/PhysRevB.77.115101}
  {\path{doi:10.1103/PhysRevB.77.115101}}.
\newline\urlprefix\url{http://link.aps.org/doi/10.1103/PhysRevB.77.115101}

\bibitem{21}
N.~G. Chernorukov, A.~V. Knyazev, Z.~S. Makarycheva,
  \href{http://link.springer.com/article/10.1134/S1066362208030016}{Synthesis,
  structure, and properties of compounds of the general formula $Ba_2A^{II}UO_6$ ($A^{II}=$ $Mn, Fe, Co, Ni, Cu, Zn, Cd, Pb$)}, Radiochemistry
  50~(3) (2008) 225--230.
\newblock \href {http://dx.doi.org/10.1134/S1066362208030016}
  {\path{doi:10.1134/S1066362208030016}}.
\newline\urlprefix\url{http://link.springer.com/article/10.1134/S1066362208030016}

\bibitem{22}
M.~W. Lufaso, P.~W. Barnes, P.~M. Woodward,
  \href{http://scripts.iucr.org/cgi-bin/paper?S010876810600262X}{Structure
  prediction of ordered and disordered multiple octahedral cation perovskites
  using \textit{{SPuDS}}}, Acta Crystallographica Section B Structural Science
  62~(3) (2006) 397--410.
\newblock \href {http://dx.doi.org/10.1107/S010876810600262X}
  {\path{doi:10.1107/S010876810600262X}}.
\newline\urlprefix\url{http://scripts.iucr.org/cgi-bin/paper?S010876810600262X}

\bibitem{23}
M.~W. Lufaso, P.~M. Woodward,
  \href{http://scripts.iucr.org/cgi-bin/paper?br0106}{Prediction of the crystal
  structures of perovskites using the software program {SPuDS}}, Acta
  Crystallographica Section B: Structural Science 57~(6) (2001) 725--738.
\newblock \href {http://dx.doi.org/10.1107/S0108768101015282}
  {\path{doi:10.1107/S0108768101015282}}.
\newline\urlprefix\url{http://scripts.iucr.org/cgi-bin/paper?br0106}

\bibitem{24}
K.~Koepernik, H.~Eschrig,
  \href{http://link.aps.org/doi/10.1103/PhysRevB.59.1743}{Full-potential
  nonorthogonal local-orbital minimum-basis band-structure scheme}, Physical
  Review B 59~(3) (1999) 1743--1757.
\newblock \href {http://dx.doi.org/10.1103/PhysRevB.59.1743}
  {\path{doi:10.1103/PhysRevB.59.1743}}.
\newline\urlprefix\url{http://link.aps.org/doi/10.1103/PhysRevB.59.1743}

\bibitem{25}
I.~Opahle, K.~Koepernik, H.~Eschrig,
  \href{http://link.aps.org/doi/10.1103/PhysRevB.60.14035}{Full-potential
  band-structure calculation of iron pyrite}, Physical Review B 60~(20) (1999)
  14035--14041.
\newblock \href {http://dx.doi.org/10.1103/PhysRevB.60.14035}
  {\path{doi:10.1103/PhysRevB.60.14035}}.
\newline\urlprefix\url{http://link.aps.org/doi/10.1103/PhysRevB.60.14035}

\bibitem{26}
J.~P. Perdew, Y.~Wang,
  \href{http://link.aps.org/doi/10.1103/PhysRevB.45.13244}{Accurate and simple
  analytic representation of the electron-gas correlation energy}, Physical
  Review B 45~(23) (1992) 13244--13249.
\newblock \href {http://dx.doi.org/10.1103/PhysRevB.45.13244}
  {\path{doi:10.1103/PhysRevB.45.13244}}.
\newline\urlprefix\url{http://link.aps.org/doi/10.1103/PhysRevB.45.13244}

\bibitem{27}
V.~I. Anisimov, I.~V. Solovyev, M.~A. Korotin, M.~T. Czyżyk, G.~A. Sawatzky,
  \href{http://link.aps.org/doi/10.1103/PhysRevB.48.16929}{Density-functional
  theory and {NiO} photoemission spectra}, Physical Review B 48~(23) (1993)
  16929--16934.
\newblock \href {http://dx.doi.org/10.1103/PhysRevB.48.16929}
  {\path{doi:10.1103/PhysRevB.48.16929}}.
\newline\urlprefix\url{http://link.aps.org/doi/10.1103/PhysRevB.48.16929}




\bibitem{29}
J.~Feng, B.~Xiao, C.~L. Wan, Z.~X. Qu, Z.~C. Huang, J.~C. Chen, R.~Zhou,
  W.~Pan,
  \href{http://www.sciencedirect.com/science/article/pii/S1359645410007962}{Electronic
  structure, mechanical properties and thermal conductivity of $Ln_2Zr_2O_7$ ({Ln}
  = {La}, {Pr}, {Nd}, {Sm}, {Eu} and {Gd}) pyrochlore}, Acta Materialia 59~(4)
  (2011) 1742--1760.
\newblock \href {http://dx.doi.org/10.1016/j.actamat.2010.11.041}
  {\path{doi:10.1016/j.actamat.2010.11.041}}.
\newline\urlprefix\url{http://www.sciencedirect.com/science/article/pii/S1359645410007962}

\bibitem{30}
Y.~P. Liu, H.~R. Fuh, Z.~R. Xiao, Y.~K. Wang,
  \href{http://www.sciencedirect.com/science/article/pii/S0925838813024341}{Theoretical
  prediction of half-metallic materials in double perovskites
  $Sr_2CrCoB'O_6$ ($B'$ = $Y, La, Zr$, and $Hf$) and $Sr_2VFeB'O_6$ ($B'$ = $Zr$ and $Hf$)}, Journal of Alloys and Compounds 586 (2014) 289--294.
\newblock \href {http://dx.doi.org/10.1016/j.jallcom.2013.10.043}
  {\path{doi:10.1016/j.jallcom.2013.10.043}}.
\newline\urlprefix\url{http://www.sciencedirect.com/science/article/pii/S0925838813024341}



\bibitem{33}
M.~A. Blanco, E.~Francisco, V.~Luaña,
  \href{http://www.sciencedirect.com/science/article/pii/S0010465503005472}{{GIBBS}:
  isothermal-isobaric thermodynamics of solids from energy curves using a
  quasi-harmonic {Debye} model}, Computer Physics Communications 158~(1) (2004)
  57--72.
\newblock \href {http://dx.doi.org/10.1016/j.comphy.2003.12.001}
  {\path{doi:10.1016/j.comphy.2003.12.001}}.
\newline\urlprefix\url{http://www.sciencedirect.com/science/article/pii/S0010465503005472}

\bibitem{34}
\href{https://books.glgoo.com/books/about/Introduction_to_the_Physics_of_the_Earth.html?hl=zh-CN&id=_YOCl86XPHoC}{Introduction
  to the {Physics} of the {Earth}'s {Interior}}.
\newline\urlprefix\url{https://books.glgoo.com/books/about/Introduction_to_the_Physics_of_the_Earth.html?hl=zh-CN&id=_YOCl86XPHoC}

\bibitem{35}
P.~Debye,
  \href{http://onlinelibrary.wiley.com/doi/10.1002/andp.19123441404/abstract}{Zur
  {Theorie} der spezifischen {Wärmen}}, Annalen der Physik 344~(14) (1912)
  789--839.
\newblock \href {http://dx.doi.org/10.1002/andp.19123441404}
  {\path{doi:10.1002/andp.19123441404}}.
\newline\urlprefix\url{http://onlinelibrary.wiley.com/doi/10.1002/andp.19123441404/abstract}















\end{thebibliography}
\end{document}